# LARGE PORTFOLIO LOSSES: A DYNAMIC CONTAGION MODEL

By Paolo Dai Pra, Wolfgang J. Runggaldier,
Elena Sartori and Marco Tolotti

*University of Padova, University of Padova, University of Padova, and
Bocconi University and Scuola Normale Superiore*

Using particle system methodologies we study the propagation of financial distress in a network of firms facing credit risk. We investigate the phenomenon of a credit crisis and quantify the losses that a bank may suffer in a large credit portfolio. Applying a large deviation principle we compute the limiting distributions of the system and determine the time evolution of the credit quality indicators of the firms, deriving moreover the dynamics of a global financial health indicator. We finally describe a suitable version of the "Central Limit Theorem" useful to study large portfolio losses. Simulation results are provided as well as applications to portfolio loss distribution analysis.

**1. Introduction.**

1.1. *General aspects.* The main purpose of this paper is to describe propagation of *financial distress* in a network of firms linked by business relationships. Once the model for *financial contagion* has been described, we quantify the impact of contagion on the losses suffered by a financial institution holding a large portfolio with positions issued by the firms.

A firm experiencing financial distress may affect the credit quality of business partners (via direct contagion) as well as of firms in the same sector (due to an information effect).

We refer to direct contagion when the actors on the market are linked by some direct partner relationship (e.g., firms in a borrowing-lending network). Reduced-form models for direct contagion can be found—among others—in Jarrow and Yu [27] for counterparty risk, Davis and Lo [13] for infectious default, Kiyotaki and Moore [28], where a model of credit chain obligations leading to default cascade is considered and Giesecke and Weber [23] for a particle system approach. Concerning the banking sector, a microeconomic liquidity equilibrium is analyzed by Allen and Gale [1].

Information effects are considered in *information-driven default models*; here the idea is that the probability of default of each obligor is influenced by a "not perfectly" observable macroeconomic variable, sometimes also referred to as *frailty*. This dependence increases the correlation between the default events. For further discussions on this point see Schönbucher [33] as well as Duffie et al. [16] and Collin-Dufresne et al. [7].

1.2. *Purpose and modeling aspects.* We propose in this paper a direct contagion model which is constructed in a general modeling framework where information effects could also be included. In addition to modeling contagion, with the approach that we shall develop we intend also to find a way to explain what is usually referred to as the *clustering of defaults* (or credit crises), meaning that there is evidence—looking at real data—of periods in which many firms end up in financial distress in a short time. A standard methodology to reproduce this real-world effect is to rely on macroeconomic factors as indicators of business cycles. These factor models seem to explain a large part of the variability of the default rates. What these models do not explain is above all clustering: as Jarrow and Yu in [27] argue, "*A default intensity that depends linearly on a set of smoothly varying macroeconomic variables is unlikely to account for the clustering of defaults around an economic recession.*"

A second issue that we would like to capture is—in some sense—more "fundamental" and refers to the nature of a credit crisis. We shall propose a model where the general "health" of the system is described by *endogenous* financial indicators, endogenous in the sense that its dynamics depends on the evolution of the variables of the system. Our aim is to show how a credit crisis can be described as a "microeconomic" phenomenon, driven by the propagation of the financial distress through the obligors.

Our model is to be considered within the class of reduced-form models and is based on *interacting intensities*. The probability of having a default somewhere in the network depends also on the state of the other obligors. The first papers on interacting intensities appear to be those by Jarrow and Yu [27], and Davis and Lo [13] on infectious default.

In our perspective the idea of a network where agents interact leads naturally to the literature of *particle systems* used in statistical mechanics. This point of view is quite new in the world of financial mathematics especially when dealing with credit risk management. Among some very recent papers we would like to mention the works by Giesecke and Weber [23], and [24] for an interacting particle approach, the papers by Frey and Backhaus [19] on credit derivatives pricing and Horst [26] on cascade processes. More developed is the use of particle and dynamical systems in the literature on financial market modeling. It has been shown that some of these models have "thermodynamic limits" that exhibit similar features compared to the limiting distributions (in particular when looking at the tails) of market returns time series. For a discussion on financial market modeling see the survey by Cont [9] and the paper by Föllmer [18] that contains an inspiring discussion on interacting agents.

Another reason to focus on particle systems is that they allow to study a credit crisis as a microeconomic phenomenon and so provide the means to explain phenomena such as default clustering that are difficult to explain by other means. In fact, interacting particle systems may exhibit what is called *phase transition* in the sense that in the limit, when the number $N$ of particles goes to infinity, the dynamics may have multiple stable equilibria. The effects of phase transition for the system with finite $N$ can be seen on different time-scales. On a long time-scale we expect to observe what is usually meant by metastability in statistical mechanics: the system may spend a very long time in a small region of the state space around a stable equilibrium of the limiting dynamics and then switch relatively quickly to another region around a different stable equilibrium. This switch, of which the rigorous analysis will be postponed to future work, occurs on a time-scale proportional to $e^{kN}$ for a suitable $k > 0$, that could be unrealistic for financial applications. The model we propose exhibits, however, a different feature that can be interpreted as a credit crisis. For certain values of the initial condition the system is driven toward a symmetric equilibrium, in which half of the firms are in good financial health. After a certain time that depends on the initial state, the system is "captured" by an unstable direction of this symmetric equilibrium, and moves toward a stable asymmetric equilibrium; during the transition to the asymmetric equilibrium, the volatility of the system increases sharply, before decaying to a stationary value. All this occurs at a time-scale of order $O(1)$ (i.e., the time-scale does not depend on $N$).

1.3. *Financial application.* As already mentioned in Section 1.1, the applied financial aim of this paper is to quantify the impact of contagion on the losses suffered by a financial institution holding a large portfolio with positions issued by the firms. In particular, we aim at obtaining a dynamic description of a risky portfolio in the context of our contagion model. The standard literature on risk management usually focuses on static models allowing to compute the distribution of a risky portfolio over a given fixed time-horizon $T$. For a recent paper that introduces a discussion relating to static and dynamic models see Dembo, Deuschel and Duffie [14].

We shall consider large homogeneous portfolios. Attention to large homogeneous portfolios becomes crucial when looking at portfolios with many small entries. Suppose a bank is holding a credit portfolio with $N = 10{,}000$ open positions with small firms; it is quite costly to simulate the dynamics of each single firm, taking into account all business ties. If the firms are supposed to be exchangeable, in the sense that the losses that they may cause to the bank in case of financial distress depend on the single firm only via its financial state indicator, it is worth evaluating a homogeneous model where $N$ goes to infinity and then to look for "large-$N$" approximations. This apparently restrictive assumption may be easily relaxed by considering many *homogeneous groups* within the network (in this context see also [19]).

We shall provide formulas to compute quantiles of the probability of excess losses in the context of our contagion model; we shall in fact determine the entire portfolio loss distribution. Other credit risk related quantities can also be computed, as we shall briefly mention at the end of Section 4.

We conclude this section by noticing that in recent years the challenging issue of describing the time evolution of the loss process connected with portfolios of many obligors has received more and more attention. Applications can be found, for example, in the literature dealing with pricing and hedging of risky derivatives such as CDOs, namely Collateralized Debt Obligations (see, e.g., the papers by Frey and Backhaus [20], Giesecke and Goldberg [22] and Schönbucher [34]).

We believe that our paper may be considered as an original contribution to the modeling of portfolio loss dynamics: to our knowledge, this is the first attempt to apply large deviations on path spaces (i.e., in a dynamic fashion) for finance or credit management purposes. For a survey on existing large deviations methods applied to finance and credit risk see Pham [31].

1.4. *Methodology.* Our interacting particle system, which describes the firms in the network, will be Markovian, but nonreversible. Usually, when the dynamics admit a reversible distribution, this distribution can be found explicitly by the detailed balance condition [see (6) below]. In the model we propose in this paper, and that will be introduced in Section 2, no reversible distribution exists. This makes it difficult to find an explicit formula for the stationary distribution. For this reason we have not pursued the "static" approach consisting in studying the $N \to +\infty$ asymptotics of the stationary distribution. We shall rather proceed in a way that in addition allows to obtain nonequilibrium properties of the system dynamics. First we study the $N \to \infty$ limiting distributions on the path space. To this effect we shall derive an appropriate *law of large numbers* based on a large deviations principle. We then study the possible equilibria of the limiting dynamics. This study leads to considering different domains of attraction corresponding to each of the *stable equilibria*. Finally, we study the *finite volume approximations* (for finite but large $N$) of the limiting distribution via a suitable version of the *Central Limit Theorem* that allows to analyze the fluctuations around this limit. As a consequence of the different domains of attraction of the limiting dynamics one obtains for finite $N$ and on ordinary time-scales an interesting behavior of the system that has an equally interesting financial interpretation, which was already alluded to at the end of Section 1.2. This behavior will also be documented by simulation results.

Our interaction model is characterized by two parameters indicating the strength of the interactions. Phase transition occurs in an open subset of the parameter space, whose boundary is a smooth curve (*critical curve*) that we determine explicitly. We shall derive the Central Limit Theorem in a fixed time-interval $[0,T]$ for every value of the parameters. We do not consider in this paper the Central Limit Theorem in the case when the time-horizon $T$ depends on $N$ itself; it will be dealt with elsewhere. When $T$ grows with $N$ we expect the behavior to depend more strongly on the parameters. In the case when the parameters belong to the uniqueness region (the complement of the closure of the region where phase transition occurs) we believe that the Central Limit Theorem should be uniform in time, while in the phase transition region the Central Limit Theorem should extend to any time-scale strictly smaller than the metastability scale (which grows exponentially in $N$). On the critical curve one expects a critical time-scale (of order $\sqrt{N}$) at which large and non-Gaussian fluctuations are observed.

For real applications the interaction parameters have to be calibrated to market data. In this paper we do not consider the issue of calibration but rather present some simulation results of the loss behavior for different values of the parameters.

The outline of the paper is as follows. The more detailed description of the model will be given in Section 2. Section 3 is devoted to stating the main limit theorems on the stochastic dynamics, in particular a law of large numbers and a central limit theorem. The financial application, in particular to large portfolio losses with specific examples, will be described in Section 4. Section 5 contains the proofs of the results stated in Sections 3 and 4. A Conclusions section completes the paper.

## 2. The model.

2.1. *A mean-field model.* In this section we describe a mean-field interaction model. What characterizes a mean-field model—within the large class of particle systems—is the absence of a "geometry" in the configuration space, meaning that each particle interacts with all the others in the same way. This "homogeneity" assumption is clearly rather restrictive; nevertheless this kind of framework has been proposed by authors in different fields. Among the others we quote Frey and Backhaus [19] for a credit risk model and Brock and Durlauf [4] for their contribution to the Social Interaction models. These models are used to capture the interaction of agents when facing any kind of decision problems. As pointed out in [19], if we are considering a large group of firms belonging to the same sector (e.g., the energy sector), then the ability of generating cash flows and the capacity of raising capital from financial institutions may be considered as "homogeneous" characteristics within the group (and this assumption is quite common in practice); we moreover recall that the final aim of this work is to study aggregate quantities for a large economy such as the expected global health of the system and large portfolio losses as well as related quantities. These considerations allow us to avoid the (costly) operation of modeling a fully heterogeneous set of firms.

Other approaches, different from the mean-field one, have also been proposed in the literature: Giesecke and Weber have chosen a local-interaction model (the Voter model[1]) assuming that each particle interacts with a fixed number $d$ of neighbors; it may be argued that the hypothesis that each firm has the same (constant) number of partners is rather unrealistic. Cont and Bouchaud (see [10]) suggest a *random graph approach*, meaning that the connections are randomly generated with some distribution functions.

The philosophy behind our model can be summarized as follows:

- We introduce only a small number of variables that, however, have a simple economic interpretation.
- We define *dynamic* rules that describe interaction between the variables.
- We keep the model as simple as possible; in particular, as we shall see, we define it in such a way that it has some symmetry properties. On one hand this may make the model less adherent to reality; on the other it leads to exact computations and still allows to show what basic features of the model produce phenomena such as clustering of defaults, phase transition, etc. More generally, it allows to show how, contrary to most models relying on macroeconomic factors, the "health" of the system can here be described by endogenous financial indicators so that a credit crisis can be viewed as a microeconomic phenomenon.

Consider a network of $N$ firms. The state of each firm is identified by two variables, that will be denoted by $\sigma$ and $\omega$ [$(\sigma_i, \omega_i)$ is the state of the $i$th firm]. The variable $\sigma$ may be interpreted as the *rating class indicator*: a low value reflects a bad rating class, that is, a higher probability of not being able to pay back obligations. The variable $\omega$ represents a more fundamental indicator of the financial health of the firm and is typically not directly observable. It could, for example, be a *liquidity indicator* as in Giesecke and Weber [23] or the *sign of the cash balances* as in Çetin et al. [5]. The important fact is that, while there is usually a strong interaction between $\sigma_i$ and $\omega_i$, the nonobservability of $\omega$ makes it reasonable to assume that $\omega_i$ cannot *directly* influence the rating indicators $\sigma_j$ for $j \neq i$.

---

[1]The Voter model assumes—roughly speaking—that the variable $\sigma_i \in \{-1, 1\}$ is more likely to take a positive value if the majority of the nearest neighbors of $i$ are in a positive state and vice versa.

In this paper we assume that the two indicators $\sigma_i, \omega_i$ can only take two values, that we label by 1 ("good" financial state) and $-1$ (financial distress). In the case of portfolios consisting of defaultable bonds, we may then refer to the rating class corresponding to $\sigma = -1$ also as "speculative grade" and that corresponding to $\sigma = +1$ as "investment grade." Although the restriction to only two possible values may appear to be unrealistic, we believe that many aspects of the qualitative behavior of the system do not really depend on this choice. On the other hand, modulo having more complex formulae, the results below can be easily extended to the case when these variables take an arbitrary finite number of values.

In our binary variable model we are naturally led to an interacting intensity model, where we have to specify the *intensities* or *rates* (inverse of the average waiting times) at which the transitions $\sigma_i \mapsto -\sigma_i$ and $\omega_i \mapsto -\omega_i$ take place. If we neglect direct interactions between the $\omega_i$'s, and we make the *mean-field* assumption that the interaction between different firms only depends on the value of the *global financial health indicator*

$$m_N^\sigma := \frac{1}{N} \sum_{i=1}^N \sigma_i,$$

we are led to consider intensities of the form

(1)
$$\sigma_i \mapsto -\sigma_i \quad \text{with intensity } a(\sigma_i, \omega_i, m_N^\sigma),$$
$$\omega_i \mapsto -\omega_i \quad \text{with intensity } b(\sigma_i, \omega_i, m_N^\sigma),$$

where $a(\cdot, \cdot, \cdot)$ and $b(\cdot, \cdot, \cdot)$ are given functions. Since both financial health and distress tend to propagate, we assume that $a(-1, \omega_i, m_N^\sigma)$ is increasing in both $\omega_i$ and $m_N^\sigma$, and $a(1, \omega_i, m_N^\sigma)$ is decreasing. Similarly, $b(\sigma_i, -1, m_N^\sigma)$ and $b(\sigma_i, 1, m_N^\sigma)$ should be respectively increasing and decreasing in their variables.

The next simplifying assumption is that the intensity $a(\sigma_i, \omega_i, m_N^\sigma)$ is actually independent of $m_N^\sigma$, that is, of the form $a(\sigma_i, \omega_i)$. Although this assumption amounts to a rather mild computational simplification, it allows to show that aggregate behavior (phase transition, etc.) may occur even in absence of a direct interaction between rating indicators.

Although a model of this generality could be fully analyzed, we make the following choice of the intensities, inspired by spin-glass systems, to make the model depend on only a few parameters:

$$\sigma_i \mapsto -\sigma_i \quad \text{with intensity } e^{-\beta\sigma_i\omega_i},$$

(2)

$$\omega_i \mapsto -\omega_i \quad \text{with intensity } e^{-\gamma\omega_i m_N^\sigma}.$$

Here $\beta$ and $\gamma$ are positive parameters which indicate the strength of the corresponding interaction. Put differently, we are considering a continuous-time Markov chain on $\{-1,1\}^{2N}$ with the following infinitesimal generator:

$$(3) \quad Lf(\underline{\sigma},\underline{\omega}) = \sum_{i=1}^N e^{-\beta\sigma_i\omega_i}\nabla_i^\sigma f(\underline{\sigma},\underline{\omega}) + \sum_{j=1}^N e^{-\gamma\omega_j m_N^\sigma}\nabla_j^\omega f(\underline{\sigma},\underline{\omega}),$$

where $\nabla_i^\sigma f(\underline{\sigma},\underline{\omega}) = f(\underline{\sigma}^i,\underline{\omega}) - f(\underline{\sigma},\underline{\omega})$ (analogously for $\nabla_i^\omega$), and where the $j$th component of $\underline{\sigma}^i$ is

$$\sigma_j^i = \begin{cases} \sigma_j, & \text{for } j \neq i, \\ -\sigma_i, & \text{for } j = i. \end{cases}$$

The rest of the paper is devoted to a detailed analysis of the above model. We conclude this subsection with some general remarks on the model we have just defined.

REMARK 2.1.

- We have viewed the variable $\sigma$ as a rating class indicator. Contrary to the standard models for rating class transitions, our rating indicator $\sigma$ is not Markov by itself, but it is Markov only if paired with $\omega$. This property is in line with empirical data and with recent research in the field of credit migration models. It is in fact well documented that real data of credit migration between rating classes exhibit a "non-Markovian" behavior. For a discussion on this topic see, for example, Christensen et al. [6]. In that paper the authors propose a *hidden* Markov process to model credit migration. The basic criticism to Markovianity is the fact that the probability of being downgraded is higher for firms that have been just downgraded. In order to capture this issue, the authors consider an "*excited*" rating state (e.g., $B^*$ from which there is a higher probability to be downgraded compared to the standard state $B$). This point of view is not far from ours, even though the mechanism of the transition is different. The downgrade to $\sigma = -1$ is higher when $(\sigma = 1, \omega = -1)$ compared to $(\sigma = 1, \omega = 1)$.
- In our model, unlike other rating class models, we do not introduce a *default* state for firms; it could be identified as a value for the pair $(\sigma,\omega)$ for which the corresponding intensities are identically zero, that is, $a(\sigma,\omega,m_N^\sigma) = b(\sigma,\omega,m_N^\sigma) = 0$ for *all* values of $m_N^\sigma$. This would have the effect of introducing a "trap state" for the system, changing drastically the long-time behavior. Even in case of defaultable firms, however, our model could be meaningful up to a time-scale in which the fraction of defaulted firms is small.

- With a choice of the intensities as in (2) we introduce a form of symmetry in our model, whereby the values $\sigma = -1$ and $\sigma = +1$ for the rating indicator turn out to be equally likely. One could, however, modify the model in order to make the value $\sigma = -1$ less (more) likely than the value $\sigma = +1$ and this could, for example, be achieved by letting the intensity for $\omega_i$ be of the form $e^{\omega_i \phi(m_N^\sigma)}$, where $\phi$ is an increasing, nonlinear and noneven function. A possible "prototype" choice would be $\phi(x) = \gamma(x - K)^+ + \delta$ with $\gamma, \delta > 0$ and $K \in (0,1)$. Note that with this latter choice we have $\phi \geq 0$ so that the value $\omega_i = +1$ (and hence also $\sigma_i = +1$) becomes more likely. Such an asymmetric setup might be more realistic in financial applications but, besides leading to more complicated derivations, it depends also on the specific application at hand. Since, as already mentioned, we want to study a model that is as simple as possible and yet capable of producing the basic features of interest, in this paper we concentrate on the "symmetric choice" in (2). The large deviation approach to the Law of Large Numbers developed in Sections 3.1 and 3.2 can be adapted to the asymmetric setup (see Remark 3.5) with no essential difference. On the other hand, our proof of the Central Limit Theorem in Section 3.3 may require more regularity on the function $\phi$ above. We leave this point for further investigation.

2.2. *Invariant measures and nonreversibility.* Mean-field models as the one we propose in this paper have already appeared, mostly in the statistical mechanics literature (see in particular [12] and [8], from which we borrow many of the mathematical tools). However, unlike what happens for the models in the cited references, we now show that our model is nonreversible. This implies that an explicit formula for the stationary distribution and its $N \to \infty$ asymptotics is not available. It is thus appropriate to follow a more specifically dynamic approach to understand the long-time behavior of the system. As already mentioned, we shall thus first study the $N \to \infty$ limit of the dynamics of the system, obtaining limit evolution equations. Then we study the equilibria of these equations. This is not necessarily equivalent to studying the $N \to \infty$ properties of the stationary distribution $\mu_N$. However, as we shall show later in this paper, this provides rather sharp information on how the system behaves for $t$ and $N$ large.

The operator $L$ given in (3) defines an irreducible, finite-state Markov chain. It follows that the process admits a unique stationary distribution $\mu_N$, that is, a distribution such that, for each function $f$ on the configuration space of $(\underline{\sigma}, \underline{\omega})$,

(4) $$\sum_{\underline{\sigma}, \underline{\omega}} \mu_N(\underline{\sigma}, \underline{\omega}) L f(\underline{\sigma}, \underline{\omega}) = 0.$$

This distribution reflects the long-time behavior of the system, in the sense that, for each $f$ and any initial distribution,
$$\lim_{t\to+\infty} E[f(\underline{\sigma}(t),\underline{\omega}(t))] = \sum_{\underline{\sigma},\underline{\omega}} \mu_N(\underline{\sigma},\underline{\omega})f(\underline{\sigma},\underline{\omega}).$$

The stationarity condition (4) is equivalent to

(5)
$$\sum_{i=1}^{N}[\mu_N(\underline{\sigma}^i,\underline{\omega})e^{\beta\sigma_i\omega_i} - \mu_N(\underline{\sigma},\underline{\omega})e^{-\beta\sigma_i\omega_i}]$$
$$+ \sum_{i=1}^{N}[\mu_N(\underline{\sigma},\underline{\omega}^i)e^{\gamma\omega_i m_N^\sigma} - \mu_N(\underline{\sigma},\underline{\omega})e^{-\gamma\omega_i m_N^\sigma}] = 0$$

for every $\underline{\sigma},\underline{\omega} \in \{-1,1\}^N$.

Simpler sufficient conditions for stationarity are the so-called *detailed balance* conditions. We say that a probability $\nu$ on $\{-1,1\}^{2N}$ satisfies the detailed balance condition for the generator $L$ if

(6)
$$\nu(\underline{\sigma}^i,\underline{\omega})e^{\beta\sigma_i\omega_i} = \nu(\underline{\sigma},\underline{\omega})e^{-\beta\sigma_i\omega_i} \quad \text{and}$$
$$\nu(\underline{\sigma},\underline{\omega}^i)e^{\gamma\omega_i m_N^\sigma} = \nu(\underline{\sigma},\underline{\omega})e^{-\gamma\omega_i m_N^\sigma}$$

for every $\underline{\sigma},\underline{\omega}$. When the detailed balance conditions (6) hold, we say the system is reversible: the stationary Markov chain with generator $L$ and marginal law $\nu$ has a distribution which is left invariant by *time-reversal*. In the case (6) admits a solution, they usually allow to derive the stationary distribution explicitly. This is not the case in our model. We have in fact:

PROPOSITION 2.2. *The detailed balance equations (6) admit no solution, except at most for one specific value of $N$.*

PROOF. By way of contradiction, assume a solution $\nu$ of (6) exists. Then one easily obtains
$$\nabla_i^\sigma \log \nu(\underline{\sigma},\underline{\omega}) = -2\beta\sigma_i\omega_i,$$
$$\nabla_i^\omega \log \nu(\underline{\sigma},\underline{\omega}) = -2\gamma\omega_i m_N^\sigma,$$

which implies
$$\nabla_i^\omega \nabla_i^\sigma \log \nu(\underline{\sigma},\underline{\omega}) = 4\beta\sigma_i\omega_i,$$
$$\nabla_i^\sigma \nabla_i^\omega \log \nu(\underline{\sigma},\underline{\omega}) = 4N^{-1}\gamma\omega_i\sigma_i.$$

This is not possible since $\nabla_i^\omega \nabla_i^\sigma \log \nu(\underline{\sigma},\underline{\omega}) \equiv \nabla_i^\sigma \nabla_i^\omega \log \nu(\underline{\sigma},\underline{\omega})$.  □

**3. Main results: law of large numbers and Central Limit Theorem.** In this section we state the results concerning the dynamics of the system $(\sigma_i[0,T], \omega_i[0,T])_{i=1}^N$ in the limit as $N \to \infty$. Note that for each value of $N$ we are considering a Markov process with generator (3). Thus, it would be more accurate to denote by $(\sigma_i^{(N)}[0,T], \omega_i^{(N)}[0,T])$ the trajectories of the variables related to the $i$th firm in the system with $N$ firms. For convenience, we consider a *fixed* probability space $(\Omega, \mathcal{F}, P)$ where all $\mathcal{D}([0,T])$-valued processes $\sigma_i^{(N)}[0,T]$, $\omega_i^{(N)}[0,T]$ are defined, and the following conditions are satisfied:

- for each $N \geq 1$ the processes $(\sigma_i^{(N)}[0,T], \omega_i^{(N)}[0,T])_{i=1}^N$ are Markov processes with infinitesimal generator (3);
- for each $N \geq 1$ the $\{-1,1\}^2$-valued random variables $(\sigma_i^{(N)}(0), \omega_i^{(N)}(0))_{i=1}^N$ are independent and identically distributed with an assigned law $\lambda$.

This last assumption on the initial distribution is stronger than what we actually need to prove the results below; however, it allows to avoid some technical aspects in the proof, that we consider not essential for the purposes of the paper. The other point, concerning the fact of realizing all processes in the same probability space, is not a restriction; we are not making any assumption on the dependence of processes with different values of $N$, so this joint realization is always possible. Its main purpose is to allow to state a *strong law of large numbers*.

Our approach proceeds according to the following three steps, to which correspond the three subsections below, namely:

  (i) look for the limit dynamics of the system $(N \to \infty)$;
  (ii) study the equilibria of the limiting dynamics;
  (iii) describe the "finite volume approximations" (for large but finite $N$) via a central limit-type result.

3.1. *Deterministic limit: large deviations and law of large numbers.* In what follows $\mathcal{D}([0,T])$ denotes the space of right-continuous, piecewise constant functions $[0,T] \to \{-1,1\}$, endowed with the Skorohod topology (see [17]). Let $(\sigma_i[0,T], \omega_i[0,T])_{i=1}^N \in \mathcal{D}([0,T])^{2N}$ denote a path of the process in the time-interval $[0,T]$ for a generic $T > 0$. If $f(\sigma_i[0,T], \omega_i[0,T])$ is a function of the trajectory of the variables related to a single firm, one is interested in the asymptotic behavior of empirical averages of the form

$$\frac{1}{N} \sum_{i=1}^N f(\sigma_i[0,T], \omega_i[0,T]) =: \int f \, d\rho_N,$$

where $\rho_N$ is the sequence of empirical measures

$$\rho_N = \frac{1}{N} \sum_{i=1}^N \delta_{(\sigma_i[0,T], \omega_i[0,T])}.$$

We may think of $\rho_N$ as a (random) element of $\mathcal{M}_1(\mathcal{D}([0,T]) \times \mathcal{D}([0,T]))$, the space of probability measures on $\mathcal{D}([0,T]) \times \mathcal{D}([0,T])$ endowed with the weak convergence topology.

Our first aim is to determine the limit of $\int f \, d\rho_N$ as $N \to \infty$, for $f$ continuous and bounded; in other words we look for the weak limit $\lim_N \rho_N$ in $\mathcal{M}_1(\mathcal{D}([0,T]) \times \mathcal{D}([0,T]))$. This corresponds to a law of large numbers with the limit being a deterministic measure. This limit, being an element of $\mathcal{M}_1(\mathcal{D}([0,T]) \times \mathcal{D}([0,T]))$, can be viewed as a stochastic process, and represents the dynamics of the system in the limit $N \to \infty$. The fluctuations of $\rho_N$ around this deterministic limit will be studied in Section 3.3 below, and this turns out to be particularly relevant in the risk analysis of a portfolio (Section 4).

The result we actually prove is a *large deviation principle*, which is much stronger than a law of large numbers. We start with some preliminary notions letting, in what follows, $W \in \mathcal{M}_1(\mathcal{D}([0,T]) \times \mathcal{D}([0,T]))$ denote the law of the $\{-1,1\}^2$-valued process $(\sigma(t), \omega(t))$ such that $(\sigma(0), \omega(0))$ has distribution $\lambda$, and both $\sigma(\cdot)$ and $\omega(\cdot)$ change sign with constant intensity 1. For $Q \in \mathcal{M}_1(\mathcal{D}([0,T]) \times \mathcal{D}([0,T]))$ let

$$H(Q|W) := \begin{cases} \int dQ \log \frac{dQ}{dW}, & \text{if } Q \ll W \text{ and } \log \frac{dQ}{dW} \in L^1(Q), \\ +\infty, & \text{otherwise}, \end{cases}$$

denote the *relative entropy* between $Q$ and $W$. Moreover, $\Pi_t Q$ denotes the marginal law of $Q$ at time $t$, and

$$\gamma_t^Q := \gamma \int \sigma \Pi_t Q(d\sigma, d\tau).$$

For a given path $(\sigma[0,T], \omega[0,T]) \in \mathcal{D}([0,T]) \times \mathcal{D}([0,T])$, let $N_t^\sigma$ (resp. $N_t^\omega$) be the process counting the jumps of $\sigma(\cdot)$ [resp. $\omega(\cdot)$]. Define

$$\begin{aligned}
F(Q) = \int \bigg[ &\int_0^T (1 - e^{-\beta \sigma(t)\omega(t)}) \, dt + \int_0^T (1 - e^{-\omega(t)\gamma_t^Q}) \, dt \\
&+ \beta \int_0^T \sigma(t) \omega(t^-) \, dN_t^\sigma + \int_0^T \omega(t) \gamma_{t^-}^Q \, dN_t^\omega \bigg] dQ,
\end{aligned} \tag{7}$$

whenever

$$\int (N_T^\sigma + N_T^\omega) \, dQ < +\infty,$$

and $F(Q) = 0$ otherwise. Finally let

$$I(Q) := H(Q|W) - F(Q).$$

We remark that, if $\int (N_T^\sigma + N_T^\omega) \, dQ = +\infty$, then $H(Q|W) = +\infty$ (this will be shown in Section 5, Lemma 5.4) and thus also $I(Q) = +\infty$.

PROPOSITION 3.1. *For each $Q \in \mathcal{M}_1(\mathcal{D}([0,T]) \times \mathcal{D}([0,T]))$, $I(Q) \geq 0$, and $I(\cdot)$ is a lower-semicontinuous function with compact level-sets [i.e., for each $k > 0$ one has that $\{Q : I(Q) \leq k\}$ is compact in the weak topology]. Moreover, for $A, C \subseteq \mathcal{M}_1(\mathcal{D}([0,T]) \times \mathcal{D}([0,T]))$ respectively open and closed for the weak topology, we have*

$$\liminf_N \frac{1}{N} \log P(\rho_N \in A) \geq - \inf_{Q \in A} I(Q), \tag{8}$$

$$\limsup_N \frac{1}{N} \log P(\rho_N \in C) \leq - \inf_{Q \in C} I(Q). \tag{9}$$

*This means that the distributions of $\rho_N$ obey a* large deviation principle (LDP) *with rate function $I(\cdot)$ (see, e.g., [15] for the definition and fundamental facts on LDP).*

The proof of Proposition 3.1 is given in Section 5 and follows from arguments similar to those in [12]. Various technical difficulties are due to unboundedness and noncontinuity of $F$, which are related to the nonreversibility of the model.

The key step to derive a law of large numbers from Proposition 3.1 is given in the following result, whose proof is also given in Section 5. In what follows, for $q \in \mathcal{M}_1(\{-1,1\}^2)$ a probability on $\{-1,1\}^2$, we define

$$m_q^\sigma := \sum_{\sigma, \omega = \pm 1} \sigma q(\sigma, \omega),$$

that can be interpreted as the expected rating under $q$.

PROPOSITION 3.2. *The equation $I(Q) = 0$ has a unique solution $Q^*$. Moreover, if $q_t \in \mathcal{M}_1(\{-1,1\}^2)$ denotes the marginal distribution of $Q^*$ at time $t$, then $q_t$ is the unique solution of the nonlinear (McKean–Vlasov) equation*

$$\frac{\partial q_t}{\partial t} = \mathcal{L} q_t, \qquad t \in [0,T], \tag{10}$$

$$q_0 = \lambda,$$

*where*

$$\mathcal{L} q(\sigma, \omega) = \nabla^\sigma [e^{-\beta \sigma \omega} q(\sigma, \omega)] + \nabla^\omega [e^{-\gamma \omega m_q^\sigma} q(\sigma, \omega)] \tag{11}$$

*with $(\sigma, \omega) \in \{-1, 1\}^2$.*

From Propositions 3.1 and 3.2, it is easy to derive the following *strong law of large numbers*.

THEOREM 3.3. *Let $Q^* \in \mathcal{M}_1(\mathcal{D}([0,T]) \times \mathcal{D}([0,T]))$ be the probability given in Proposition 3.2. Then*

$$\rho_N \to Q^* \quad \text{almost surely}$$

*in the weak topology.*

PROOF. Let $Q^*$ be the unique zero of the rate function $I(\cdot)$ as given by Proposition 3.2. Let $B_{Q^*}$ be an arbitrary open neighborhood of $Q^*$ in the weak topology. By the upper bound in Proposition 3.1, we have

$$\limsup_N \frac{1}{N} \log P(\rho_N \notin B_{Q^*}) \leq -\inf_{Q \notin B_{Q^*}} I(Q) < 0,$$

where the last inequality comes from lower semicontinuity of $I(\cdot)$, compactness of its level sets and the fact that $I(Q) > 0$ for every $Q \neq Q^*$. Indeed, if $\inf_{Q \notin B_{Q^*}} I(Q) = 0$, then there exists a sequence $Q_n \notin B_{Q^*}$ such that $I(Q_n) \to 0$. By the compactness of the level sets there exists then a subsequence $Q_{n_k} \to \bar{Q} \notin B_{Q^*}$. By lower semicontinuity it then follows $I(\bar{Q}) \leq \liminf I(Q_{n_k}) = 0$ which contradicts $I(Q) > 0$ for $q \neq Q^*$. By the above inequality we thus have that $P(\rho_N \notin B_{Q^*})$ decays to 0 exponentially fast. By a standard application of the Borel–Cantelli lemma, we obtain that $\rho_n \to Q^*$ almost surely. □

3.2. *Equilibria of the limiting dynamics: phase transition.* Equation (10) describes the dynamics of the system with generator (3) in the limit as $N \to +\infty$. In this section we determine the equilibrium points, or stationary (in $t$) solutions of (10), that is, solutions of $\mathcal{L}q_t = 0$ and, more generally, the large time behavior of its solutions. First of all, it is convenient to reparametrize the unknown $q_t$ in (10).

Let $q$ be a probability on $\{-1,1\}^2$. Note that each $f:\{-1,1\}^2 \to \mathbb{R}$ can be written in the form $f(\sigma,\omega) = a\sigma + b\omega + c\sigma\omega + d$. It follows that $q$ is completely identified by the expectations

(12)
$$m_\mu^\sigma := \sum_{\sigma,\omega=\pm 1} \sigma q(\sigma,\omega),$$

$$m_\mu^\omega := \sum_{\sigma,\omega=\pm 1} \omega q(\sigma,\omega),$$

$$m_\mu^{\sigma\omega} := \sum_{\sigma,\omega=\pm 1} \sigma\omega q(\sigma,\omega).$$

In particular, if $q = q_t$, the marginal of $Q^*$ appearing in Proposition 3.2, then we write $m_t^\sigma$ for $m_{q_t}^\sigma$, and similarly for $m_t^\omega, m_t^{\sigma\omega}$. In order to rewrite (10) in terms of the new variables $m_t^\sigma, m_t^\omega, m_t^{\sigma\omega}$, observe that

$$\dot{m}^\sigma = \sum_{\sigma,\omega=\pm 1} \sigma \dot{q}_t(\sigma,\omega) = \sum_{\sigma,\omega=\pm 1} \sigma \mathcal{L}q_t.$$

On the other hand, a straightforward computation shows that, for every probability $q$,

$$\sum_{\sigma,\omega=\pm 1} \sigma \mathcal{L}q = 2\sinh(\beta)m_q^\omega - 2\cosh(\beta)m_q^\sigma,$$

giving

$$\dot{m}_t^\sigma = 2\sinh(\beta)m_t^\omega - 2\cosh(\beta)m_t^\sigma.$$

By making similar computations for $m_t^\omega, m_t^{\sigma\omega}$, it is shown that (10) can be rewritten in the following form:

$$\begin{aligned}
\dot{m}_t^\sigma &= 2\sinh(\beta)m_t^\omega - 2\cosh(\beta)m_t^\sigma, \\
(13) \quad \dot{m}_t^\omega &= 2\sinh(\gamma m_t^\sigma) - 2\cosh(\gamma m_t^\sigma)m_t^\omega, \\
\dot{m}_t^{\sigma\omega} &= 2\sinh(\beta) + 2\sinh(\gamma m_t^\sigma)m_t^\sigma - 2(\cosh(\beta) + \cosh(\gamma m_t^\sigma))m_t^{\sigma\omega},
\end{aligned}$$

with initial condition $m_0^\sigma = m_\lambda^\sigma$, $m_0^{\sigma\omega} = m_\lambda^{\sigma\omega}$, $m_0^\omega = m_\lambda^\omega$. Note that $m_t^{\sigma\omega}$ does not appear in the first and in the second equation in (13); this means that the differential system (13) is essentially two-dimensional: first one solves the two-dimensional system (on $[-1,1]^2$)

$$(14) \qquad (\dot{m}_t^\sigma, \dot{m}_t^\omega) = V(m_t^\sigma, m_t^\omega),$$

with $V(x,y) = (2\sinh(\beta)y - 2\cosh(\beta)x, 2\sinh(\gamma x) - 2y\cosh(\gamma x))$, and then one solves the third equation in (13), which is linear in $m_t^{\sigma\omega}$. Note also that to any $(m_*^\sigma, m_*^\omega)$ satisfying $V(m_*^\sigma, m_*^\omega) = 0$, there corresponds a unique $m_*^{\sigma\omega} := \frac{\sinh(\beta) + m_*^\sigma \sinh(\gamma m_*^\sigma)}{\cosh(\beta) + \cosh(\gamma m_*^\sigma)}$ such that $(m_*^\sigma, m_*^\omega, m_*^{\sigma\omega})$ is an equilibrium (stable solution) of (13). Moreover, if $m_t^\sigma \to m_*^\sigma$ as $t \to +\infty$, then $m_t^{\sigma\omega} \to m_*^{\sigma\omega}$. Thus, to discuss the equilibria of (13) and their stability, it is enough to analyze (14) and for this we have the following proposition, where by "linearly stable equilibrium" we mean a pair $(\bar{x}, \bar{y})$ such that $V(\bar{x}, \bar{y}) = 0$, and the linearized system $(\dot{x}, \dot{y}) = DV(\bar{x}, \bar{y})(x - \bar{x}, y - \bar{y})$ is stable, that is, the eigenvalues of the Jacobian matrix $DV(\bar{x}, \bar{y})$ have all negative real parts.

THEOREM 3.4. (i) *Suppose* $\gamma \leq \frac{1}{\tanh(\beta)}$. *Then (14) has $(0,0)$ as a unique equilibrium solution, which is globally asymptotically stable, that is, for every initial condition* $(m_0^\sigma, m_0^\omega)$, *we have*

$$\lim_{t \to +\infty} (m_t^\sigma, m_t^\omega) = (0,0).$$

(ii) *For* $\gamma < \frac{1}{\tanh(\beta)}$ *the equilibrium* $(0,0)$ *is linearly stable. For* $\gamma = \frac{1}{\tanh(\beta)}$ *the linearized system has a neutral direction, that is, $DV(0,0)$ has one zero eigenvalue.*

(iii) *For $\gamma > \frac{1}{\tanh(\beta)}$ the point $(0,0)$ is still an equilibrium for (14), but it is a saddle point for the linearized system, that is, the matrix $DV(0,0)$ has two nonzero real eigenvalues of opposite sign. Moreover (14) has two linearly stable solutions $(m_*^\sigma, m_*^\omega)$, $(-m_*^\sigma, -m_*^\omega)$, where $m_*^\sigma$ is the unique strictly positive solution of the equation*

$$\text{(15)} \qquad x = \tanh(\beta)\tanh(\gamma x),$$

*and*

$$\text{(16)} \qquad m_*^\omega = \frac{1}{\tanh(\beta)} m_*^\sigma.$$

(iv) *For $\gamma > \frac{1}{\tanh(\beta)}$, the phase space $[-1,1]^2$ is bipartitioned by a smooth curve $\Gamma$ containing $(0,0)$ such that $[-1,1]^2 \setminus \Gamma$ is the union of two disjoint sets $\Gamma^+, \Gamma^-$ that are open in the induced topology of $[-1,1]^2$. Moreover*

$$\lim_{t \to +\infty}(m_t^\sigma, m_t^\omega) = \begin{cases} (m_*^\sigma, m_*^\omega), & \text{if } (m_0^\sigma, m_0^\omega) \in \Gamma^+, \\ (-m_*^\sigma, -m_*^\omega), & \text{if } (m_0^\sigma, m_0^\omega) \in \Gamma^-, \\ (0,0), & \text{if } (m_0^\sigma, m_0^\omega) \in \Gamma. \end{cases}$$

PROOF. See Section 5. □

REMARK 3.5. The results in this section are specific to our model with the symmetry properties as induced by the specification of the intensities in (2). With an asymmetric setup such as described in Remark 2.1, (15) becomes

$$x = \tanh(\beta)\tanh(\phi(x))$$

thus allowing more flexibility in the position of the equilibria. In particular, by letting $\phi(x) = \gamma(x - K)^+ + \delta$, while still having three equilibria, we may choose their relative position by suitably choosing the values for $\gamma, K, \delta$. Notice that in this way we also increase the number of parameters in our model.

3.3. *Analysis of fluctuations: Central Limit Theorem.* Having established a law of large numbers $\rho_N \to Q^*$, it is natural to analyze fluctuations around the limit, that is, the rate at which $\rho_N$ converges to $Q^*$ and the asymptotic distribution of $\rho_N - Q^*$.

To study the asymptotic distribution of $\rho_N - Q^*$ there are at least the following two possible approaches:

(i) An approach based on a functional central limit theorem using a result in [2] that relates large deviations with the Central Limit Theorem (see [35], Chapter 3, for some results in this direction).

(ii) A weak convergence-type approach based on uniform convergence of the generators (see [17]).

In this paper we shall follow an approach of the second type; more precisely we shall provide a dynamical interpretation of the law of large numbers discussed in Theorem 3.3. Let $\psi:\{-1,1\}^2 \to \mathbb{R}$, and define $\rho_N(t)$ by

$$\int \psi \, d\rho_N(t) := \frac{1}{N} \sum_{i=1}^{N} \psi(\sigma_i(t), \omega_i(t)).$$

In other words, $\rho_N(t)$ is the marginal of $\rho_N$ at time $t$ and we also have $m_N^\sigma(t) = m_{\rho_N(t)}^\sigma$. Note that, for each fixed $t$, $\rho_N(t)$ is a probability on $\{-1,1\}^2$, and so, by the considerations leading to (12), it can be viewed as a three-dimensional object. Thus $(\rho_N(t))_{t \in [0,T]}$ is a three-dimensional flow. A simple consequence of Theorem 3.3 is the following convergence of flows:

(17) $$(\rho_N(t))_{t \in [0,T]} \to (q_t)_{t \in [0,T]} \quad \text{a.s.,}$$

where the convergence of flows is meant in the uniform topology. Since the flow of marginals contains less information than the full measure of paths, the law of large numbers in (17) is weaker than the one in Theorem 3.3. However, the corresponding fluctuation flow

$$(\sqrt{N}(\rho_N(t) - q_t))_{t \in [0,T]}$$

is also a finite-dimensional flow, and it allows for a very explicit characterization of the limiting distribution. The following theorem gives the asymptotic behavior of this fluctuation flow; its proof is given in Section 5.

THEOREM 3.6. *Consider the following three-dimensional fluctuation process:*

$$x_N(t) := \sqrt{N}(m_{\rho_N(t)}^\sigma - m_t^\sigma),$$
$$y_N(t) := \sqrt{N}(m_{\rho_N(t)}^\omega - m_t^\omega),$$
$$z_N(t) := \sqrt{N}(m_{\rho_N(t)}^{\sigma\omega} - m_t^{\sigma\omega}).$$

*Then $(x_N(t), y_N(t), z_N(t))$ converges as $N \to \infty$, in the sense of weak convergence of stochastic processes, to a limiting three-dimensional Gaussian process $(x(t), y(t), z(t))$ which is the unique solution of the following linear stochastic differential equation:*

(18) $$\begin{pmatrix} dx(t) \\ dy(t) \\ dz(t) \end{pmatrix} = A(t) \begin{pmatrix} x(t) \\ y(t) \\ z(t) \end{pmatrix} dt + D(t) \begin{pmatrix} dB_1(t) \\ dB_2(t) \\ dB_3(t) \end{pmatrix}$$

where $B_1, B_2, B_3$ are independent, standard Brownian motions,

$$A(t) = 2 \begin{pmatrix} -\cosh(\beta) & \sinh(\beta) & 0 \\ -\gamma m_t^\omega \sinh(\gamma m_t^\sigma) + \gamma \cosh(\gamma m_t^\sigma) & -\cosh(\gamma m_t^\sigma) & 0 \\ \sinh(\gamma m_t^\sigma) + \gamma m_t^\sigma \cosh(\gamma m_t^\sigma) - \gamma m_t^{\sigma\omega} \sinh(\gamma m_t^\sigma) & 0 & -(\cosh(\beta) + \cosh(\gamma m_t^\sigma)) \end{pmatrix},$$

$$\frac{D(t)D^*(t)}{2} = \begin{pmatrix} -m_t^{\sigma\omega} \sinh(\beta) + \cosh(\beta) & 0 & -m_t^\sigma \sinh(\beta) + m_t^\omega \cosh(\beta) \\ 0 & -m_t^\omega \sinh(\gamma m_t^\sigma) + \cosh(\gamma m_t^\sigma) & m_t^\sigma \cosh(\gamma m_t^\sigma) - m_t^{\sigma\omega} \sinh(\gamma m_t^\sigma) \\ -m_t^\sigma \sinh(\beta) + m_t^\omega \cosh(\beta) & m_t^\sigma \cosh(\gamma m_t^\sigma) - m_t^{\sigma\omega} \sinh(\gamma m_t^\sigma) & -m_t^{\sigma\omega} \sinh(\beta) + \cosh(\beta) - m^\omega \sinh(\gamma m_t^\sigma) + \cosh(\gamma m_t^\sigma) \end{pmatrix},$$

and $(x(0), y(0), z(0))$ have a centered Gaussian distribution with covariance matrix

$$(19) \quad \begin{pmatrix} 1 - (m_\lambda^\sigma)^2 & m_\lambda^{\sigma\omega} - m_\lambda^\sigma m_\lambda^\omega & m_\lambda^\omega - m_\lambda^\sigma m_\lambda^{\sigma\omega} \\ m_\lambda^{\sigma\omega} - m_\lambda^\sigma m_\lambda^\omega & 1 - (m_\lambda^\omega)^2 & m_\lambda^\sigma - m_\lambda^{\sigma\omega} m_\lambda^\omega \\ m_\lambda^\omega - m_\lambda^\sigma m_\lambda^{\sigma\omega} & m_\lambda^\sigma - m_\lambda^{\sigma\omega} m_\lambda^\omega & 1 - (m_\lambda^{\sigma\omega})^2 \end{pmatrix}.$$

Theorem 3.6 guarantees that, for each $t > 0$, the distribution of $(x_N(t), y_N(t), z_N(t))$ is asymptotically Gaussian, and provides a method to compute the limiting covariance matrix. Indeed, denote by $\Sigma_t$ the covariance matrix of $(x(t), y(t), z(t))$. A simple application of Itô's rule to (18) shows that $\Sigma_t$ solves the Lyapunov equation

$$(20) \quad \frac{d\Sigma_t}{dt} = A(t)\Sigma_t + \Sigma_t A(t)^* + D(t)D^*(t).$$

In order to solve (20), it is convenient to interpret $\Sigma$ as a vector in $\mathbb{R}^{3 \times 3} = \mathbb{R}^3 \otimes \mathbb{R}^3$. To avoid ambiguities, for a $3 \times 3$ matrix $C$ we write $\mathrm{vec}(C)$ whenever we interpret it as a vector. It is easy to check that (20) can be rewritten as follows

$$(21) \quad \frac{d(\mathrm{vec}(\Sigma_t))}{dt} = (A(t) \otimes I + I \otimes A(t))\mathrm{vec}(\Sigma_t) + \mathrm{vec}(D(t)D^*(t)),$$

where "$\otimes$" denotes the tensor product of matrices. Equation (21) is linear, so its solution can be given an explicit expression and can be computed after having solved (13). More importantly, the behavior of $\Sigma_t$ for large $t$ can be obtained explicitly as follows.

A. *Case* $\gamma < \frac{1}{\tanh(\beta)}$. In this case we have shown in Theorem 3.4 that the solution $(m_t^\sigma, m_t^\omega, m_t^{\sigma\omega})$ of (13) converges to $(0, 0, \tanh(\beta))$ as $t \to +\infty$. In particular, one immediately obtains the limits

(22) $$A := \lim_{t \to +\infty} A(t), \qquad DD^* := \lim_{t \to +\infty} D(t)D^*(t).$$

A direct inspection (see the Appendix) shows that $A$ has three real strictly negative eigenvalues. Moreover, the eigenvalues of the matrix $A \times I + I \times A$ are all of the form $\lambda_i + \lambda_j$ where $\lambda_i$ and $\lambda_j$ are eigenvalues of $A$, and therefore they are all strictly negative. It follows from (21) that $\lim_{t \to +\infty} \Sigma_t = \Sigma$ where

(23) $$\operatorname{vec}(\Sigma) = -(A \otimes I + I \otimes A)^{-1} \operatorname{vec}(DD^*).$$

B. *Case* $\gamma > \frac{1}{\tanh(\beta)}$. Also in this case, by Theorem 3.4, the limit

$$\lim_{t \to +\infty} (m_t^\sigma, m_t^\omega, m_t^{\sigma\omega})$$

exists. Disregarding the exceptional case in which the initial condition of (13) belongs to the stable manifold $\Gamma$ introduced in Theorem 3.4(iv), the limit above equals either $(m_*^\sigma, m_*^\omega, m_*^{\sigma\omega})$, or $(-m_*^\sigma, -m_*^\omega, m_*^{\sigma\omega})$, depending on the initial condition, where $(m_*^\sigma, m_*^\omega, m_*^{\sigma\omega})$ are obtained by Theorem 3.4(iii). In both cases one obtains as in (22) the limits $A$ and $DD^*$, and we show in the Appendix that also in this case the eigenvalues of $A$ are real and strictly negative, so that $\lim_{t \to +\infty} \Sigma_t = \Sigma$ is obtained as in (23).

C. *Case* $\gamma = \frac{1}{\tanh(\beta)}$. In this case, as shown in the Appendix, the limiting matrix $A$ is singular; it follows that the limit $\lim_{t \to +\infty} \Sigma_t$ does not exist, as one eigenvalue of $\Sigma_t$ grows polynomially in $t$. This means that, for *critical* values of the parameters, the size of normal fluctuations around the deterministic limit grows in time. Similarly to what is done in [8] for reversible models, it is possible to determine the critical long-time behavior of the fluctuation by a suitable space–time scaling in the model, giving rise to nonnormal fluctuations. More precisely, one can show the following convergence in distribution:

$$N^{1/4}(m_{\rho_N}^\cdot(\sqrt{N}t) - m^\cdot(\sqrt{N}t)) \stackrel{N \to \infty}{\Longrightarrow} Z$$

where $Z$ is non-Gaussian. This result is contained in [32].

We now state an immediate corollary of Theorem 3.6 concerning the fluctuations of the global health indicator; this will be used in the next section on large portfolio losses.

COROLLARY 3.7. *As $N \to \infty$ we have that*

$$\sqrt{N}[m^\sigma_{\rho_N(t)} - m^\sigma_t]$$

*converges in law to a centered Gaussian random variable $Z$ with variance*

(24) $$V(t) = \Sigma_{11}(t),$$

*where $\Sigma(t)$ solves (20) and $m^\sigma_t$ solves (13).*

We conclude this section with the following:

REMARK 3.8. The evolution equation (20) for the covariance matrix $\Sigma_t$ is coupled with the McKean–Vlasov equation (13), and their joint behavior exhibits interesting aspects even before the system gets close to the stable fixed point. In particular, in the case $\gamma > \frac{1}{\tanh(\beta)}$, if the initial condition is sufficiently close to the stable manifold $\Gamma$, the system (13) spends some time close to the symmetric equilibrium $(0,0)$ before drifting to one of the stable equilibria. A closer look at (20) shows that when the system is close to the neutral equilibrium, the covariance matrix $\Sigma$ grows exponentially fast in time, causing sharp peaks in the variances. This is related to the credit crisis mentioned in the Introduction. A more detailed discussion on this point is given in the next section, in relation with applications to portfolio losses.

**4. Portfolio losses.** We address now the problem of computing losses in a portfolio of positions issued by the $N$ firms. A rather general modeling framework is to consider the total loss that a bank may suffer due to a risky portfolio at time $t$ as a random variable defined by $L^N(t) = \sum_i L_i(t)$. Different specifications for the single (marginal) losses $L_i(t)$ can be chosen accounting for heterogeneity, time dependence, interaction, macroeconomic factors and so on. A punctual treatment of this general modeling framework can be found in the book by McNeil, Frey and Embrechts [29]. For a comparison with the most widely used industry examples of credit risk models see Frey and McNeil [21], Crouhy, Galai and Mark [11] or Gordy [25]. The same modeling insights are also developed in the most recent literature on risk management and large portfolio losses analysis; see [14, 19, 23, 26] for different specifications.

In this paper we adopt the point of view of Giesecke and Weber [23]. The idea is to compute the aggregate losses as a sum of marginal losses $L_i(t)$, of which the distribution is supposed to depend on the realization of the variable $\sigma_i$, that is, on the rating class. In particular, conditioned on the realization of $\sigma$, the marginal losses will be assumed to be independent and identically distributed (the independence condition can be weakened; see

Example 4.4 below). More precisely, we assume given a suitable conditional distribution function $G_x$, $x \in \{-1, 1\}$, namely

$$
(25) \qquad G_x(u) := P(L_i(t) \leq u | \sigma_i(t) = x)
$$

where the first and second moments are well defined, namely

$$
(26) \qquad l_1 := E(L_i(t) | \sigma_i(t) = 1) < E(L_i(t) | \sigma_i(t) = -1) =: l_{-1}
$$

and

$$
(27) \qquad v_1 := Var(L_i(t) | \sigma_i(t) = 1), \qquad v_{-1} := Var(L_i(t) | \sigma_i(t) = -1).
$$

The inequality in (26) specifies that we expect to lose more when in financial distress.

The aggregate loss of a portfolio of volume $N$ at time $t$ is then defined as

$$
L^N(t) = \sum_{i=1}^{N} L_i(t).
$$

We recall the definition of the global health indicators $m_N^\sigma(t) := \frac{1}{N} \sum_{i=1}^{N} \sigma_i(t)$, and $m_t^\sigma := \int \sigma \, dq_t$ where $q_t$ solves the McKean–Vlasov equation [see (10)].

We also introduce a deterministic time function, which will be seen to represent an "asymptotic" loss when the number of firms goes to infinity, namely

$$
(28) \qquad L(t) = \frac{(l_1 - l_{-1})}{2} m_t^\sigma + \frac{(l_1 + l_{-1})}{2}.
$$

We state now the main result of this section.

THEOREM 4.1. *Assume $L_i(t)$ has a distribution of the form (25). Then for $t \in [0, T]$ with generic $T > 0$ and for any value of the parameters $\beta > 0$ and $\gamma > 0$, we have*

$$
\sqrt{N} \left( \frac{L^N(t)}{N} - L(t) \right) \to Y \sim N(0, \hat{V}(t))
$$

*in distribution, where $L(t)$ has been defined in (28) and*

$$
(29) \qquad \hat{V}(t) = \frac{(l_1 - l_{-1})^2 V(t)}{4} + \frac{(1 + m_t^\sigma) v_1}{2} + \frac{(1 - m_t^\sigma) v_{-1}}{2},
$$

*with $V(t)$ as defined in (24).*

PROOF. See Section 5. □

REMARK 4.2. The Gaussian approximation in Theorem 4.1 leads in particular to

$$P(L^N(t) \geq \alpha) \approx \mathcal{N}\left(\frac{NL(t) - \alpha}{\sqrt{N}\sqrt{\hat{V}(t)}}\right). \tag{30}$$

By the symmetry of the model, the above Gaussian approximation for the losses is appropriate for a wide (depending on $N$) range of values of $\alpha$. If we modify the model to become asymmetric as discussed in Remark 2.1 and, more precisely, we modify it so that $\sigma = -1$ becomes much less likely than $\sigma = +1$, then for a "realistic" value of $N$, the number of firms with $\sigma_i = -1$ could be too small for the Gaussian approximation to be sufficiently precise. One could then rather consider a Poisson-type approximation instead.

We shall now provide examples illustrating possible specifications for the marginal loss distributions where, without loss of generality, we assume a unitary loss (e.g., loss due to a corporate bond) when a firm is in the bad state.

We start with a very basic example where we assume that the marginal losses (when conditioned on the value of $\sigma$) are deterministic. This means that the riskiness of the loss portfolio is related only to the number of firms in financial distress and so we can use directly the results of Section 3, in particular of Corollary 3.7.

EXAMPLE 4.3. Suppose that marginal losses are described as follows:

$$L_i(t) = \begin{cases} 1, & \text{if } \sigma_i(t) = -1, \\ 0, & \text{if } \sigma_i(t) = 1. \end{cases}$$

On the other hand

$$L^N(t) = \sum_{i=1}^{N} \frac{1 - \sigma_i(t)}{2}.$$

Recalling that $m_N^\sigma(t) = \frac{1}{N}\sum_i \sigma_i(t)$, by Corollary 3.7 [see also (30)], we can compute various risk measures related to the portfolio losses such as the following Var-type measure:

$$P(L^N(t) \geq \alpha) = P\left(\frac{N - Nm_N^\sigma(t)}{2} \geq \alpha\right) = P\left(m_N^\sigma(t) \leq \frac{N - 2\alpha}{N}\right)$$
$$\approx \mathcal{N}\left(\frac{-2\alpha + (1 - m_t^\sigma)N}{\sqrt{N}\sqrt{V(t)}}\right) = \mathcal{N}\left(\frac{-2\alpha + 2L^\infty(t)N}{\sqrt{N}\sqrt{V(t)}}\right), \tag{31}$$

where $L^\infty(t) := \lim_{N\to\infty} \frac{L^N(t)}{N} = \lim_{N\to\infty} \sum_i \frac{1-\sigma_i(t)}{2N} = \frac{1-m_t^\sigma}{2}$.

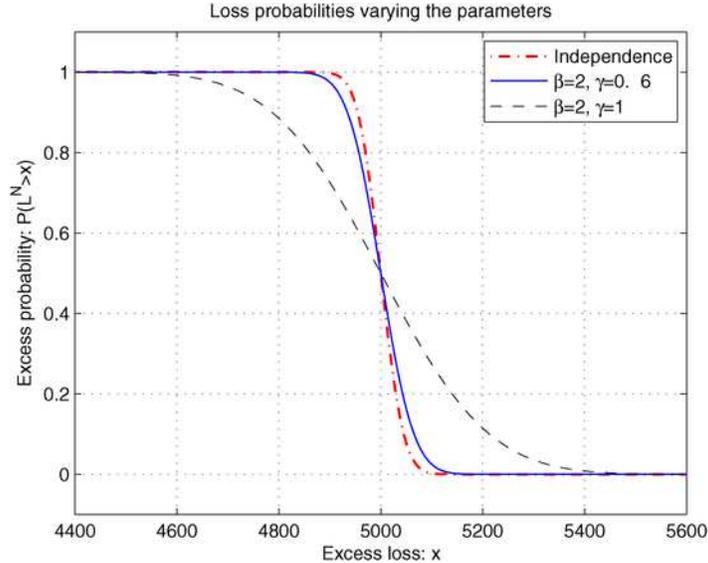

FIG. 1. *Excess loss in a large portfolio ($N = 10{,}000$) for different values of the parameters $\gamma$ and $\beta$ compared with the independence case.*

Looking at a portfolio of $N = 10{,}000$ small firms, we compute the excess loss probability for different values of the parameters $\beta, \gamma$ comparing them with the benchmark case where there is no interaction at all, that is, where $\beta = \gamma = 0$ ("independence case"). In Figure 1 we show the cumulative probability of having excess losses for the same portfolios. In this figure we see that, when the dependence increases, variance and risk measures increase as well.

More general specifications are already suggested in the existing literature. For example, one could consider the losses to depend also on a random *exogenous factor* $\Psi$; more precisely, the marginal losses $L_i(t)$ are independent and identically distributed *conditionally* to the realizations of the $\sigma_i(t)$'s *and* of $\Psi$. The conditional distributions

$$G_x(u) := P(L_i(t) \le u | \sigma_i(t) = x, \Psi)$$

are random variables, as well as the corresponding moments $l_1, l_{-1}, v_1, v_{-1}$.

In particular in the following example we apply our approach to a very tractable class of models, the "Bernoulli mixture models." This kind of modeling has been used in the context of cyclical correlations, that is in models where exogenous factors are supposed to characterize the evolution of the indicator of defaults (the classical factor models). In the context of contagion-based models this class was first introduced by Giesecke and Weber in [23].

EXAMPLE 4.4 (Bernoulli mixture models). Assume that the marginal losses $L_i(t)$ are Bernoulli mixtures, that is,

$$(32) \quad L_i(t) = \begin{cases} 1, & \text{with probability } P(\sigma_i(t), \Psi), \\ 0, & \text{with probability } 1 - P(\sigma_i(t), \Psi), \end{cases}$$

where the mixing derives not only from the rating class indicator $\sigma_i(t)$ of firm $i$, but also from an exogenous factor $\Psi \in \mathbb{R}^p$ that represents macroeconomic variables that reflect the business cycle and thus allow for both contagion and cyclical effects on the rating probabilities.

Notice that, with the above specification, the quantities defined in (26) and (27) now depend on the random factor $\Psi$, that is,

$$l_1 = P(1, \Psi), \qquad v_1 = P(1, \Psi)(1 - P(1, \Psi)) \qquad \text{and analogously for } l_{-1}, v_{-1}.$$

Consequently, the asymptotic loss function $L(t)$ as well as the variance of the Gaussian approximation $\hat{V}(t)$ defined in (28) and (29) are also functions of $\Psi$. With a slight abuse of notation we shall write $L_\psi(t)$ [respectively $\hat{V}_\psi(t)$] for the asymptotic loss (variance) at time $t$ given that $\Psi = \psi$.

Next we give a possible expression for the mixing distribution for $P(\sigma, \Psi)$ that is in line with existing models on contagion. Let $a$ and $b_i$, $i = 1, 2$, be nonnegative weight factors. Let us assume for simplicity that $\Psi \in \mathbb{R}$ is a Gamma distributed random variable. Define then

$$P(\sigma, \Psi) = 1 - \exp\left\{-a\Psi - b_1\left(\frac{1-\sigma}{2}\right) - b_2\right\}.$$

This specification follows the *CreditRisk+* modeling structure, even though in the standard industry examples direct contagion is not taken into account. Notice that the factor $\frac{1-\sigma}{2}$ increases the probability of default for the firms in the bad rating class ($\sigma = -1$). Using (30) we have that

$$P(L_N(t) \geq \alpha) \approx \int \mathcal{N}\left(\frac{NL_\psi(t) - \alpha}{\sqrt{N\hat{V}_\psi(t)}}\right) df_\Psi(\psi),$$

where $f_\Psi$ is the density function of the Gamma random variable $\Psi$.

In Figure 2 we plot the excess loss probability in the case where $a = 0.1$, $b_1 = 1$, $b_2 = 0.5$ and $\beta = 1.5$ is supposed to be fixed. We compare different specifications for $\Psi$ and $\gamma$. In particular we consider the following cases:

$$\Psi = 4.5, \qquad \gamma = 0.6; \qquad \Psi = 4.5, \qquad \gamma = 1.1;$$
$$\Psi \sim \Gamma(2.25, 2), \qquad \gamma = 1.1.$$

The shape of the excess losses suggests that the loss may be sensibly higher in the case of high uncertainty about the value of the macroeconomic factor [$\Psi \sim \Gamma(2.25; 2)$] and in the case of high level of contagion ($\gamma = 1.1$). Notice that in all three situations we are in the subcritical case, since the critical value for $\gamma$ is $\gamma_c = 1/\tanh(\beta) \simeq 1.105$. This also implies that the equilibrium value is the same in the three situations and depends only on $\Psi$.

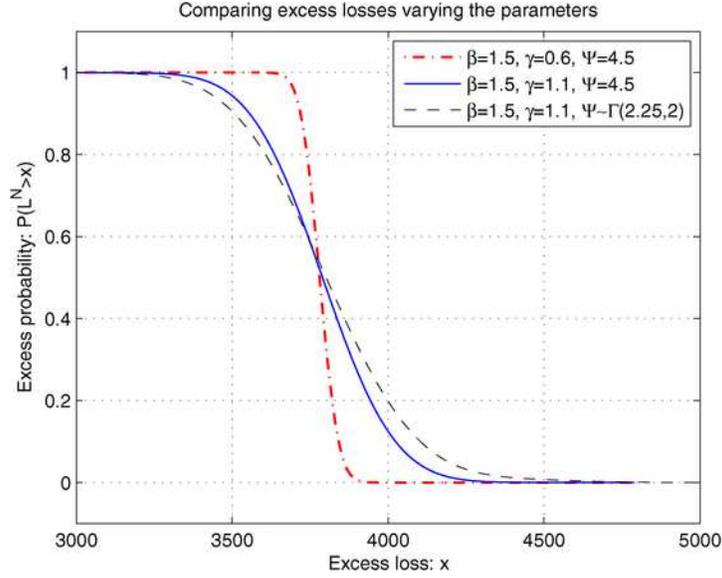

Fig. 2. *Loss amount in a large portfolio ($N = 10{,}000$) in the case of marginal losses which (depending on the rating class) are distributed as Bernoulli random variables for which the parameter depends on $\Psi$.*

REMARK 4.5. Notice that the asymptotic loss distribution in the above Bernoulli mixture model does not only depend on a mixing parameter as in standard Bernoulli mixture models but, via $L(t)$, it depends also on the value $m_t^\sigma$ of the asymptotic average global health indicator. Moreover, compared to Giesecke and Weber [23], we are able to quantify the time-varying fluctuations of the global indicator $m_{\rho_N(t)}^\sigma$. We shall see that this may sensibly influence the distribution of losses in particular when looking at two different time horizons $T_1$ and $T_2$ before and after a credit crisis.

REMARK 4.6. Further examples may be considered, in particular when the distribution of the marginal losses $L_i(t)$ depends on the entire past trajectory of the rating indicator $\sigma_i$, taking, for example, into account how long the firm has been in the bad state. Instead of depending simply on $\sigma_i(t)$, the distribution of $L_i(t)$ could then be made dependent on $S_i(t) := \mathbb{I}_{\{\int_0^t ((1-\sigma_i(s))/2)\,ds \geq \delta t\}}$ with $\delta \in (0,1)$, which is equal to 1 if firm $i$ has spent a fraction $\delta$ of time in the bad state. Corresponding to (32) we would then have

$$L_i(t) = \begin{cases} 1, & \text{with probability } P(S_i(t), \Psi), \\ 0, & \text{with probability } 1 - P(S_i(t), \Psi). \end{cases}$$

This model is not a straightforward extension of Example 4.4. In fact the theory developed above, in particular the Central Limit Theorem result in Section 3.3, does not appear to be strong enough to handle it. For this purpose an approach based on a functional central limit theorem that was alluded to at the beginning of Section 3.3 would be more appropriate. This, however, goes beyond the scope of the present paper.

Let us point out that in the examples above we have considered only the problem of computing large portfolio losses which led to examples where we computed (approximately) the quantiles $P(L^N(t) \geq \alpha)$ where $\alpha$ is a (large) integer. From here, one could then compute the probability that the loss ratio $\frac{L^N(t)}{N}$ belongs to a given interval and this would then allow to compute (approximately) for our contagion model also other quantities in a risk-sensitive environment. In any case notice that Theorem 4.1 provides the entire asymptotic distribution for the portfolio losses.

In the previous examples we have described large portfolio losses at a predetermined time horizon $T$ for different specifications of the conditional loss distribution. In what follows, we shall describe in more detail how the phenomenon of a *credit crisis* may be explained in our setting and how this issue may influence the quantification of losses. This dynamic point of view on risk management that accounts for the possibility of a credit crisis in the market, is one of the main contributions of this work.

As one could expect, the possibility of having a credit crisis is related to the existence of particular conditions on the market, more precisely to certain levels of interaction between the obligors (i.e., the parameters $\beta$ and $\gamma$) and certain values of the state variables describing the rating classes and the fundamentals (i.e., $\sigma$ and $\omega$).

4.1. *Simulation results.* To illustrate the situation we shall now present some simulation results. We shall proceed along two steps: the first one relates more specifically to the particle system, the second to the portfolio losses.

*Step 1 (Domains of attraction).* In Section 3.2 we have characterized all the equilibria of the system depending on the values of the parameters. In particular we have shown that for supercritical values, by which we mean $\gamma > \frac{1}{\tanh(\beta)}$, there are two asymmetric equilibrium configurations in the space $(m^\sigma, m^\omega)$ that, for our symmetric model, are symmetric to one another and are defined as $(m_*^\sigma, m_*^\omega)$ and $(-m_*^\sigma, -m_*^\omega)$.

In particular, Theorem 3.4 allows to characterize their *domains of attraction*, that is, the sets of initial conditions that lead the trajectory to one of the equilibria, and we shall denote them by $\Gamma^+$ and $\Gamma^-$. Numerical simulations provide diagrams as in Figure 3.

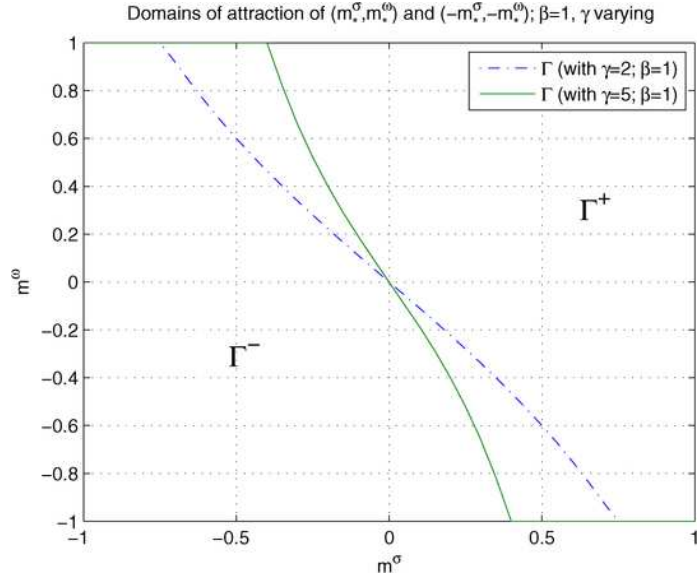

FIG. 3. *Domains of attraction $\Gamma^+$ for $(m_*^\sigma, m_*^\omega)$ and $\Gamma^-$ for $(-m_*^\sigma, -m_*^\omega)$ and their boundary $\Gamma$ for $\beta = 1$ and varying $\gamma$. Here the critical value for $\gamma$ is $\gamma_c := 1/\tanh(\beta) \simeq 1.313$.*

*Step 2 (Credit crises).* We show results from numerical simulations that detect the crises when the values of the parameters are supercritical and the initial conditions are "near" the boundary of the domains of attraction, that is, near $\Gamma$. Given the symmetry of our model, the behavior of the system will be perfectly symmetric when starting in either $\Gamma^+$ or $\Gamma^-$, but the typical credit crisis corresponds to what happens in $\Gamma^-$, so that below we shall illustrate this latter case. The analysis in an asymmetric model would be analogous.

In Figure 4 we have plotted a trajectory starting in $(m_0^\sigma, m_0^\omega) \in \Gamma^-$ but near the boundary. It can be seen that the path moves toward $(m^\sigma, m^\omega) = (0, 0)$ and then leaves it decaying to the stable equilibrium.

Concerning the time evolution, we see in Figure 5 that, for an initial condition in $\Gamma^-$ and near the boundary, the variable $m_t^\sigma$ (the same would happen also with $m_t^\omega$ that for clarity is not plotted) is first attracted to the unstable value zero, around which it spends a long time before moving to the stable equilibrium value $m_*^\sigma$. This can be explained, in financial terms, as follows:

Suppose that at the initial time the market conditions are such that $(m^\sigma, m^\omega)$ are in $\Gamma^-$ but close to the curve $\Gamma$. Then for a while the system moves close to the stable manifold $\Gamma$ toward $(0, 0)$, until it gets "captured" by the unstable direction of the equilibrium point $(0,0)$. Since the system configuration belongs to $\Gamma^-$, the new stable equilibrium that the system is attracted to is given by $(-m_*^\sigma, -m_*^\omega)$.

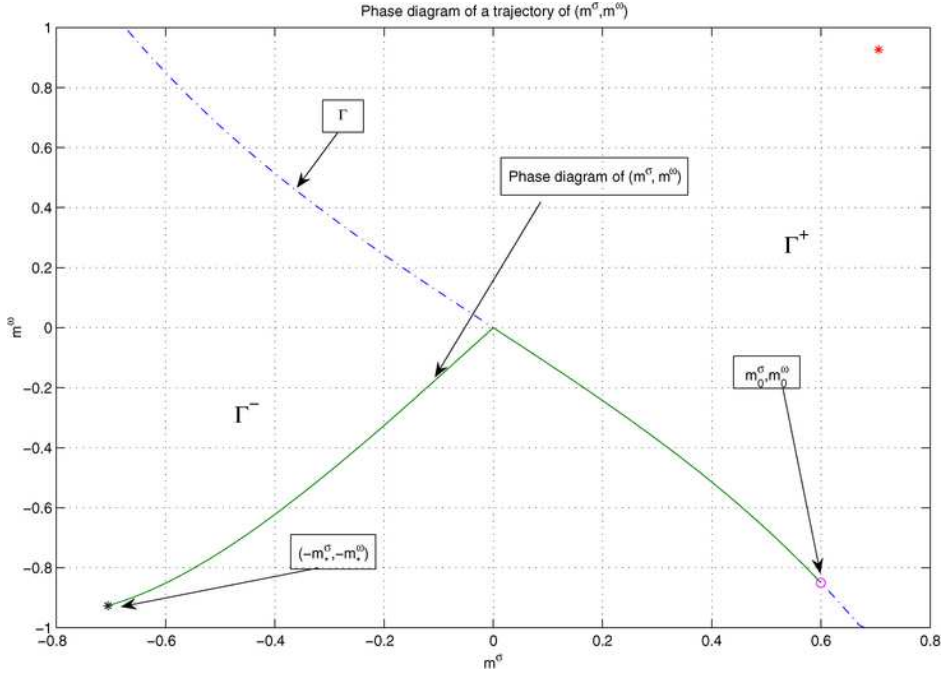

FIG. 4. *Domains of attraction $\Gamma^+$ for $(m_*^\sigma, m_*^\omega)$ and $\Gamma^-$ for $(-m_*^\sigma, -m_*^\omega)$ and phase diagram of $(m_t^\sigma, m_t^\omega)$ with initial conditions $(m_0^\sigma, m_0^\omega) = (0.6, -0.85)$ when $\beta = 1$ and $\gamma = 2.3$ [here $\gamma_c = 1/\tanh(\beta) \simeq 1.313$].*

This situation represents (in a stylized manner) what we intend as a credit crisis: the state $(0,0)$ may be considered as a "credit bubble," the decay toward the stable equilibrium mimics a credit crisis (i.e., a crash in the credit market).

As soon as the system moves away from $(0,0)$, the uncertainty (volatility) increases quickly and the credit quality indicators move to the *stable* configuration changing completely the picture of the market (the speed of the convergence depends on the level of interaction).

This situation is also well illustrated by the loss probability computed before and after the crisis (i.e., in certain time instants $T_1$ and $T_2$). In Figure 6 we see the excess probability of suffering a loss larger than $x$ for the case of Example 4.4 with an exogenous parameter $\Psi \sim \Gamma(2.25; 2)$. One can see that before the crisis both the expected loss and the variance may be underestimated as well as the corresponding risk measures. Put differently, a model that does not distinguish between stable and unstable equilibria (does not take credit crises into account) may underestimate the excess loss probability, since it does not recognize in the given situation the possibility of a sudden crash.

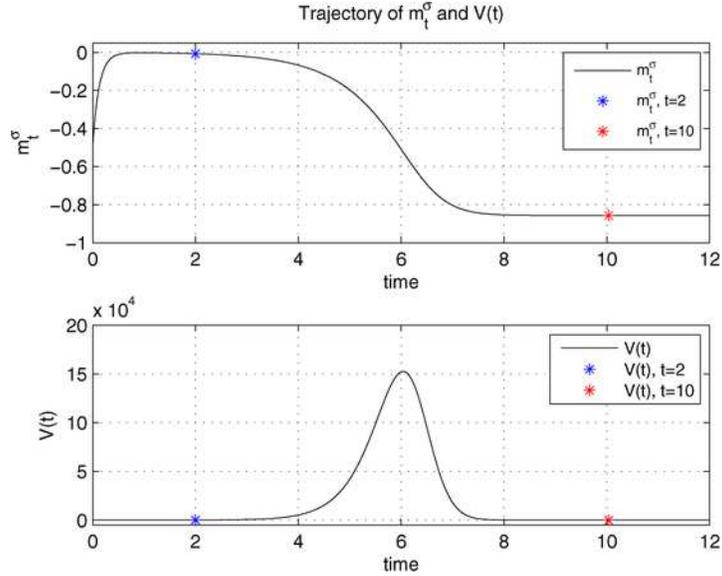

FIG. 5. *Trajectory of $m_t^\sigma$ and $V(t)$ with initial conditions $m_0^\sigma = -0.5$, $m_0^\omega = 0.395$ when $\beta = 1.5$ and $\gamma = 2.1$ [here $\gamma_c = 1/\tanh(\beta) \simeq 1.105$]. We have marked by $(*)$ the time horizons $T_1 = 2$ and $T_2 = 10$ before and after the crisis where in Figure 6 we shall compute the excess loss probabilities.*

Finally we mention the fact that for different levels of interaction we can distinguish between a smoothly varying business cycle and a crisis. When $\beta$ and $\gamma$, the parameters describing the level of interaction, are sufficiently small, the business cycle (described in our simple model by the proportion of firms in the rating classes) evolves smoothly and the induced variance (level of uncertainty about the number of bad rated firms) is lower compared to the crisis case. In Figure 7 we show this fact for two levels of $\beta$ and $\gamma$, both supercritical.

## 5. Proofs.

5.1. *Proofs of Propositions 3.1 and 3.2.* One of the main tools in this proof is the *Girsanov formula* for Markov chains. Since a Markov chain is a functional of the multivariate point process that counts the jumps between all pairs of states, this formula can be derived from the corresponding Girsanov formula for point processes (see, e.g., [3], Section 4.2). We state it here for completeness.

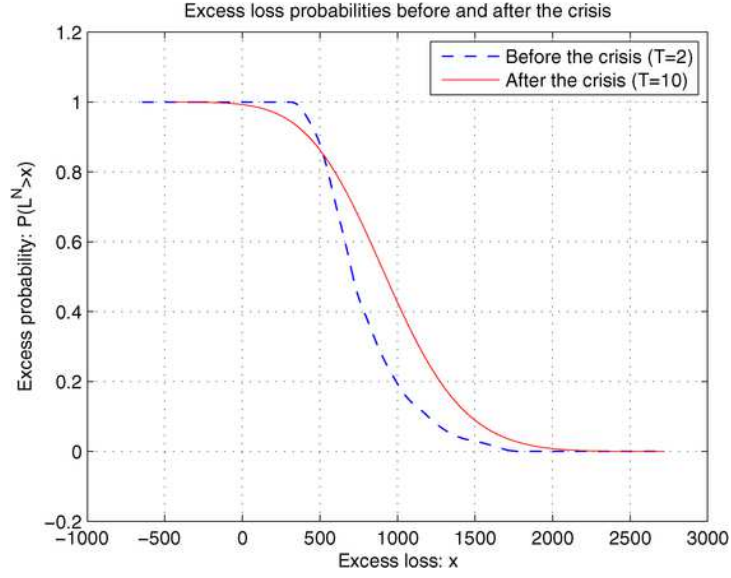

Fig. 6.  *Excess probability of losses in a portfolio of $N = 10{,}000$ obligors, $\beta = 1.5$ and $\gamma = 2.1$ computed in $T_1 = 2$ and $T_2 = 10$, namely before and after the crisis in the case of Example 4.4 with $\Psi \sim \Gamma(2.25; 2)$ [here $\gamma_c = 1/\tanh(\beta) \simeq 1.105$].*

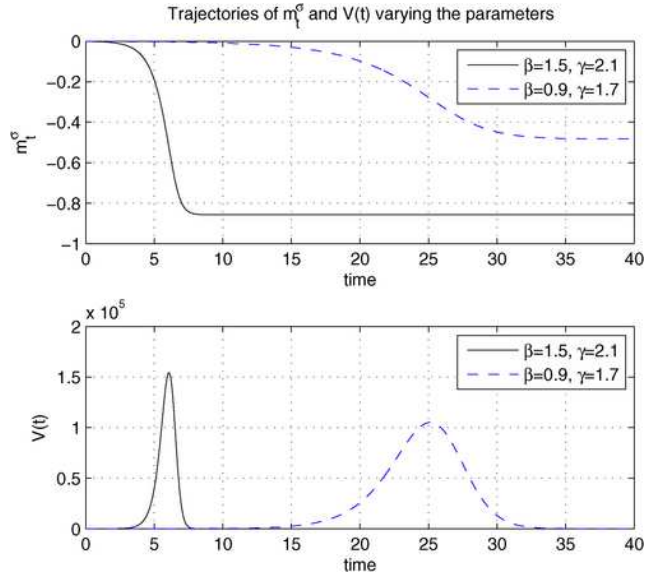

Fig. 7.  *Trajectories of $m_t^\sigma$ and $V(t)$ for different levels of interaction, that is, letting $\beta$ and $\gamma$ vary. In the case of higher values we really see a crisis and a corresponding peak in the uncertainty in the market. In the case of smaller values the number of bad rated firms decreases smoothly to a new equilibrium, that is, toward a bad business cycle. The critical values for $\gamma$ are, respectively, $1/\tanh(1.5) \simeq 1.105$ and $1/\tanh(0.9) \simeq 1.396$.*

PROPOSITION 5.1. *Let $S$ be a finite set, and $(X(t))_{t \in [0,T]}$, $(Y(t))_{t \in [0,T]}$ two $S$-valued Markov chains with infinitesimal generators, respectively,*

$$Lf(x) = \sum_{y \neq x} L_{x,y}[f(y) - f(x)],$$

$$Mf(x) = \sum_{y \neq x} M_{x,y}[f(y) - f(x)].$$

*Assume $X(0)$ and $Y(0)$ have the same distribution, and denote by $P_X$ and $P_Y$ the law of the two processes on the appropriate set of trajectories in the time-interval $[0,T]$. Assume that whenever $M_{x,y} = 0$ also $L_{x,y} = 0$. Then $P_X \ll P_Y$, and*

$$\frac{dP_X}{dP_Y}(x([0,T]))$$

$$= \exp\left[\int_0^T \sum_{y \neq x(t)} (M_{x(t),y} - L_{x(t),y}) \, dt + \int_0^T \log \frac{L_{x(t^-),x(t)}}{M_{x(t^-),x(t)}} \, dN_t\right],$$

*where $x(t^-) := \lim_{s \uparrow t} x(s)$, $\log \frac{0}{0} = 1$ and $N_t$ is the counting process that counts the jumps of the trajectory $x([0,T])$.*

In what follows we denote by $P_N$ the law on the path space of $(\underline{\sigma}[0,T], \underline{\omega}[0,T]) \in (\mathcal{D}([0,T]))^{2N}$ under the interacting dynamics, with initial conditions such that $(\sigma_i^{(N)}(0), \omega_i^{(N)}(0))_{i=1}^N$ are independent and identically distributed with an assigned law $\lambda$ (see beginning of Section 3). As in Section 3.1 we let $W \in \mathcal{M}_1(\mathcal{D}([0,T]) \times \mathcal{D}([0,T]))$ denote the law of the $\{-1,1\}^2$-valued process $(\sigma(t), \omega(t))$ such that $(\sigma(0), \omega(0))$ has distribution $\lambda$, and both $\sigma(\cdot)$ and $\omega(\cdot)$ change sign with constant rate 1. By $W^{\otimes N}$ we mean the product of $N$ copies of $W$. We begin with some preliminary lemmas.

LEMMA 5.2.

(33) $$\frac{dP_N}{dW^{\otimes N}}(\underline{\sigma}[0,T], \underline{\omega}[0,T]) = \exp[NF(\rho_N(\underline{\sigma}[0,T], \underline{\omega}[0,T]))],$$

*where $F$ is the function defined in (7).*

PROOF. Let $(N_t^\sigma(i))_{i=1}^N$ be the multivariate counting process which counts the jumps of $\sigma_i$ for $i = 1, \ldots, N$, and $(N_t^\omega(i))_{i=1}^N$ be the multivariate counting process which counts the jumps of $\omega_i$ for $i = 1, \ldots, N$. Since each jump of the trajectory $(\underline{\sigma}[0,T], \underline{\omega}[0,T])$ is counted by exactly one of the above counting processes, Proposition 5.1 applied to this case yields

$$\frac{dP_N}{dW^{\otimes N}}(\underline{\sigma}[0,T], \underline{\omega}[0,T])$$

$$= \exp\left[\sum_{i=1}^{N} \int_0^T (1 - e^{-\beta\sigma_i(t)\omega_i(t)})\, dt + \sum_{i=1}^{N} \int_0^T \log e^{-\beta\sigma_i(t^-)\omega_i(t^-)}\, dN_t^\sigma(i) \right.$$

$$+ \sum_{i=1}^{N} \int_0^T (1 - e^{-\gamma\omega_i(t)m^\sigma_{\rho_N(t)}})\, dt$$

$$\left. + \sum_{i=1}^{N} \int_0^T \log e^{-\gamma\omega_i(t^-)m^\sigma_{\rho_N(t^-)}}\, dN_t^\omega(i)\right].$$

Since, with probability 1 with respect to $W^{\otimes N}$, there are no simultaneous jumps, we have

$$\sum_{i=1}^{N} \int_0^T \log e^{-\beta\sigma_i(t^-)\omega_i(t^-)}\, dN_t^\sigma(i) = \sum_{i=1}^{N} \int_0^T -\beta(-\sigma_i(t))\omega_i(t)\, dN_t^\sigma(i)$$

and

$$\sum_{i=1}^{N} \int_0^T \log e^{-\gamma\omega_i(t^-)m^\sigma_{\rho_N(t^-)}}\, dN_t^\omega(i) = \sum_{i=1}^{N} \int_0^T -\gamma(-\omega_i(t))\, m^\sigma_{\rho_N(t)}\, dN_t^\omega(i),$$

from which (33) follows easily after having observed that, $W^{\otimes N}$ almost surely,

$$\int (N_T^\sigma + N_T^\omega)\, d\rho_N < +\infty,$$

and that simultaneous jumps of $\sigma$ and $\omega$ do not occur under $dW^{\otimes N}$. □

The main problem in the proof of Proposition 3.1 is related to the fact that the function $F$ in (7) is neither continuous nor bounded. The following technical lemmas have the purpose of circumventing this problem. In what follows, we let

(34) $$\mathcal{I} := \left\{Q \in \mathcal{M}_1(\mathcal{D}[0,T]^2) : \int (N_T^\sigma + N_T^\omega)\, dQ < +\infty\right\}.$$

We first define, for $r > 0$ and $Q \in \mathcal{I}$,

$$F_r(Q) = \int \left[\int_0^T (r - e^{-\beta\sigma(t)\omega(t)})\, dt + \int_0^T (r - e^{-\omega(t)\gamma_t^Q})\, dt\right.$$

(35)
$$+ \int_0^T (\beta\sigma(t)\omega(t^-) - \log r)\, dN_t^\sigma$$

$$\left. + \int_0^T (\omega(t)\gamma_{t^-}^Q - \log r)\, dN_t^\omega\right] dQ.$$

Note that $F = F_1$. Moreover, Lemma 5.2 can be easily extended to show that

$$\text{(36)} \quad \frac{dP_N}{dW_r^{\otimes N}}(\underline{\sigma}[0,T], \underline{\omega}[0,T]) = \exp[NF_r(\rho_N(\underline{\sigma}[0,T], \underline{\omega}[0,T]))],$$

where $W_r$ is the law of the $\{-1,1\}^2$-valued process $\sigma(t), \omega(t)$ such that $(\sigma(0), \omega(0))$ has distribution $\lambda$, and both $\sigma(\cdot)$ and $\omega(\cdot)$ change sign with constant rate $r$.

LEMMA 5.3. *For $0 < r \leq \min(e^{-\beta}, e^{-\gamma})$, $F_r$ is lower semicontinuous on $\mathcal{I}$. For $r \geq \max(e^\beta, e^\gamma)$, $F_r$ is upper semicontinuous.*

PROOF. By definition of weak topology the fact that the map

$$Q \mapsto \int \left[ \int_0^T (r - e^{-\beta\sigma(t)\omega(t)}) \, dt + \int_0^T (r - e^{-\omega(t)\gamma_t^Q}) \, dt \right] dQ$$

is continuous is rather straightforward (since $Q$-expectations of bounded continuous functions in $\mathcal{D}([0,T])$ are continuous in $Q$). Thus we only have to deal with the term

$$\text{(37)} \quad \int \left[ \int_0^T (\beta\sigma(t)\omega(t^-) - \log r) \, dN_t^\sigma \right] dQ$$

$$+ \int \left[ \int_0^T (\omega(t)\gamma_{t^-}^Q - \log r) \, dN_t^\omega \right] dQ.$$

We show that for $0 < r \leq \min(e^{-\beta}, e^{-\gamma})$ the expression in (37) is lower semicontinuous in $Q \in \mathcal{I}$. This shows that $F_r$ is lower semicontinuous. The case $r \geq \max(e^\beta, e^\gamma)$ is treated similarly.

For $\varepsilon > 0$ consider the function $\varphi_\varepsilon : \mathcal{D}[0,T] \to \mathbb{R}$ defined by

$$\varphi_\varepsilon(\eta) := \begin{cases} \dfrac{1}{\varepsilon}, & \text{if } \eta(t) \text{ jumps for some } t \in (0, \varepsilon], \\ 0, & \text{otherwise.} \end{cases}$$

Given $\eta \in \mathcal{D}([0,T])$ we define $\eta(s)$ for $s > T$ by letting $\eta(s) \equiv \eta(T)$. Then, letting $\theta_t$ denote the shift operator, we have that, for $t \in [0,T]$, $\theta_t \eta$ is the element of $\mathcal{D}([0,T])$ given by $\theta_t \eta(s) := \eta(t+s)$. Consider now two functions $f, g : \{-1,1\}^2 \to \mathbb{R}$, and define $f_\varepsilon, g_\varepsilon : \mathcal{D}[0,T]^2 \to \mathbb{R}$ by

$$f_\varepsilon(\sigma_{[0,T]}, \omega_{[0,T]}) := \inf\{f(\sigma(t), \omega(t)) : t \in (0,\varepsilon)\},$$

and similarly for $g_\varepsilon$. Then define

$$\Phi_\varepsilon(\sigma_{[0,T]}, \omega_{[0,T]}) := \int_0^T f_\varepsilon(\theta_t\sigma, \theta_t\omega)\varphi_\varepsilon(\theta_t\sigma) \, dt + \int_0^T g_\varepsilon(\theta_t\sigma, \theta_t\omega)\varphi_\varepsilon(\theta_t\omega) \, dt.$$

The key to the continuation of the proof below are the following two properties of $\Phi_\varepsilon$. These properties are essentially straightforward, and their proofs are omitted:

- $\Phi_\varepsilon$ is continuous and bounded on $\{(\sigma_{[0,T]}, \omega_{[0,T]}): N_T^\sigma + N_T^\omega < +\infty\}$.
- Suppose $f, g \geq 0$. Then, assuming $\sigma_{[0,T]}, \omega_{[0,T]}$ have a finite number of jumps, $\Phi_\varepsilon(\sigma_{[0,T]}, \omega_{[0,T]})$ increases when $\varepsilon \downarrow 0$ to

$$\int_0^T f(\sigma_{t-}, \omega_{t-}) \, dN_t^\sigma + \int_0^T g(\sigma_{t-}, \omega_{t-}) \, dN_t^\omega.$$

Therefore by monotone convergence

$$\int \left[ \int_0^T f(\sigma_{t-}, \omega_{t-}) \, dN_t^\sigma + \int_0^T g(\sigma_{t-}, \omega_{t-}) \, dN_t^\omega \right] dQ$$

$$= \sup_{\varepsilon > 0} \int \Phi_\varepsilon(\sigma_{[0,T]}, \omega_{[0,T]}) \, dQ.$$

In particular, the map

$$Q \mapsto \int \left[ \int_0^T f(\sigma_{t-}, \omega_{t-}) \, dN_t^\sigma + \int_0^T g(\sigma_{t-}, \omega_{t-}) \, dN_t^\omega \right] dQ$$

is lower semicontinuous on $\mathcal{I}$.

Now, for $r \leq \min(e^{-\beta}, e^{-\gamma})$, the function $f(\sigma, \omega) = -\beta \sigma \omega - \log r$ is nonnegative. As for the function $g$, that should be $-\omega(t)\gamma_t^Q - \log r$, we notice that it is not a function of $(\sigma, \omega)$, but rather a function of $(\sigma, \Pi_t Q)$, thus depending explicitly on $t$ and $Q$. However, due to its boundedness and the fact that $\gamma_t^Q$ is continuous in $Q$ uniformly in $t, \sigma$, the argument above applies with minor modifications thus leading to the conclusion of the proof. $\square$

LEMMA 5.4.  *Let $Q \in \mathcal{M}_1(\mathcal{D}([0,T])^2)$ be such that $H(Q|W) < +\infty$. Then $Q \in \mathcal{I}$. The same result applies if $W_r$ replaces $W$.*

PROOF.  By the entropy inequality (see (6.2.14) in [15])

$$\int N_T^\sigma \, dQ \leq \log \int e^{N_T^\sigma} \, dW + H(Q|W).$$

But $N_T^\sigma$ has Poisson distribution under $W$, so $\int e^{N_T^\sigma} \, dW < +\infty$. By applying the same argument to $N_T^\omega$, the proof is completed. This proof extends with no modifications to the case $r \neq 1$. $\square$

LEMMA 5.5.  *The function*

$$I(Q) := H(Q|W) - F(Q)$$

*is lower semicontinuous on $\mathcal{M}_1(\mathcal{D}[0,T]^2)$.*

PROOF. It is well known (see [15], Lemma 6.2.13) that the entropy $H(Q|W)$ is lower semicontinuous in $Q$ in all of $\mathcal{M}_1(\mathcal{D}([0,T])^2)$. Moreover, by definition, $F(Q) < +\infty$ for every $Q$, and so we have $H(Q|W) = I(Q)$ whenever $H(Q|W) = +\infty$. Since, by Lemma 5.4, $H(Q|W) = +\infty$ for $Q \notin \mathcal{I}$, we are left to prove the following two statements:

(i) $I(Q)$ is lower semicontinuous in $\mathcal{I}$.
(ii) If $H(Q|W) = +\infty$ and $Q_n \to Q$ weakly, then $I(Q_n) \to +\infty$.

The following key identity, which holds for $r > 0$, is a simple consequence of the definition of relative entropy and of the Girsanov formula for Markov chains.

$$
\begin{aligned}
H(Q|W_r) &= H(Q|W) + \int \log \frac{dW}{dW_r} dQ \\
&= H(Q|W) + 2T(r-1) + \log r \int (N_T^\sigma + N_T^\omega) dQ.
\end{aligned}
\tag{38}
$$

In particular, by Lemma 5.4, we have that $H(Q|W) < +\infty \iff H(Q|W_r) < +\infty$. A simple consequence of (38) is then the following:

$$I(Q) = H(Q|W_r) - F_r(Q), \tag{39}$$

where the difference in (39) is meant to be $+\infty$ whenever $H(Q|W_r) = +\infty$ [which is equivalent to $H(Q|W) = +\infty$].

We are now ready to prove (i) and (ii). To prove (i) it is enough to choose $r \geq \max(e^\beta, e^\gamma)$ and use Lemma 5.3. Moreover, for the same choice of $r$, the stochastic integrals in (35) are nonpositive, so $F_r(Q) \leq 2Tr$. Therefore, if $H(Q|W) = +\infty$ and $Q_n \to Q$,

$$\liminf I(Q_n) \geq \liminf H(Q_n|W_r) - 2Tr = +\infty,$$

where the last equality follows from lower semicontinuity of $H(\cdot|W_r)$ and $H(Q|W_r) = +\infty$. Thus (ii) is proved. □

LEMMA 5.6. *The function $I(Q)$ has compact level sets, that is, for every $k > 0$ the set $\{Q : I(Q) \leq k\}$ is compact.*

PROOF. Choosing, as above, $r \geq \max(e^\beta, e^\gamma)$, we have that $F_r(Q) \leq 2Tr$ for every $Q$. Thus, by (39),

$$\{Q : I(Q) \leq k\} \subseteq \{Q : H(Q|W_r) \leq k + 2Tr\}.$$

Since (see [15], Lemma 6.2.13) the relative entropy has compact level sets, $\{Q : I(Q) \leq k\}$ is contained in a compact set. Moreover, by lower semicontinuity of $I$, $\{Q : I(Q) \leq k\}$ is closed, and this completes the proof. □

LEMMA 5.7. *For every $r > 0$ there exists $\delta > 1$ such that*

$$\limsup_{N \to +\infty} \frac{1}{N} \log \int \exp[\delta N F_r(\rho_N)] \, dW_r^{\otimes N} < +\infty.$$

PROOF. We give the proof for $r = 1$; the modifications for the general case are obvious. The proof consists of rather simple manipulations. The idea can be summarized as follows. If $\delta = 1$, then by Lemma 5.2, $\exp[\delta N F(\rho_N)]$ is the Radon–Nikodym derivative of $P_N$ with respect to $W^{\otimes N}$, and therefore has expectation 1. For $\delta > 1$, we write $\delta F(\rho_N) = F_1(\rho_N) + F_2(\rho_N)$ in such a way that $F_2$ is bounded and $\exp[N F_1(\rho_N)]$ is a Radon–Nikodym derivative of a probability with respect to $W^{\otimes N}$. More specifically, observe that, using (7),

$$\delta N F(\rho_N) = \sum_{i=1}^{N} \int_0^T (\delta - \delta e^{-\beta \sigma_i(t) \omega_i(t)}) \, dt + \sum_{i=1}^{N} \int_0^T \delta \beta \sigma_i(t) \omega_i(t^-) \, dN_t^\sigma(i)$$

$$+ \sum_{i=1}^{N} \int_0^T (\delta - \delta e^{-\gamma \omega_i(t) m_{\rho_N(t)}^\sigma}) \, dt$$

$$+ \sum_{i=1}^{N} \int_0^T \delta \gamma \omega_i(t) m_{\rho_N(t^-)}^\sigma \, dN_t^\omega(i)$$

$$= \sum_{i=1}^{N} \int_0^T (1 - e^{-\delta \beta \sigma_i(t) \omega_i(t)}) \, dt + \sum_{i=1}^{N} \int_0^T \delta \beta \sigma_i(t) \omega_i(t) \, dN_t^\sigma(i)$$

$$+ \sum_{i=1}^{N} \int_0^T (1 - e^{-\delta \gamma \omega_i(t) m_{\rho_N(t)}^\sigma}) \, dt + \sum_{i=1}^{N} \int_0^T \delta \gamma \omega_i(t) m_{\rho_N(t)}^\sigma \, dN_t^\omega(i)$$

$$+ \sum_{i=1}^{N} \int_0^T (\delta - \delta e^{-\beta \sigma_i(t) \omega_i(t)} - (1 - e^{-\delta \beta \sigma_i(t) \omega_i(t)})) \, dt$$

$$+ \sum_{i=1}^{N} \int_0^T (\delta - \delta e^{-\gamma \omega_i(t) m_{\rho_N(t)}^\sigma} - (1 - e^{-\delta \gamma \omega_i(t) m_{\rho_N(t)}^\sigma})) \, dt$$

$$= N F_1(\rho_N) + N F_2(\rho_N),$$

where

$$N F_1(\rho_N) := \sum_{i=1}^{N} \int_0^T (1 - e^{-\delta \beta \sigma_i(t) \omega_i(t)}) \, dt + \sum_{i=1}^{N} \int_0^T \delta \beta \sigma_i(t) \, \omega_i(t) \, dN_t^\sigma(i)$$

$$+ \sum_{i=1}^{N} \int_0^T (1 - e^{-\delta \gamma \omega_i(t) m_{\rho_N(t)}^\sigma}) \, dt + \sum_{i=1}^{N} \int_0^T \delta \gamma \omega_i(t) m_{\rho_N(t)}^\sigma \, dN_t^\omega(i)$$

and

$$NF_2(\rho_N) := \sum_{i=1}^{N} \int_0^T \left(\delta - \delta e^{-\beta \sigma_i(t) \omega_i(t)} - (1 - e^{-\delta\beta\sigma_i(t)\omega_i(t)})\right) dt$$
$$+ \sum_{i=1}^{N} \int_0^T \left(\delta - \delta e^{-\gamma \omega_i(t) m^\sigma_{\rho_N}(t)} - (1 - e^{-\delta\gamma\omega_i(t) m^\sigma_{\rho_N}(t)})\right) dt.$$

Note that $\exp[NF_1(\rho_N)]$ has the same form of $\exp[NF(\rho_N)]$ after having replaced $\beta$ by $\delta\beta$. In particular, $\int \exp[NF_1(\rho_N)] dW^{\otimes N} = 1$. Moreover, it is easy to see that

$$F_2(\rho_N) \leq T(2\delta - \delta(e^{-\beta} + e^{-\gamma}) - 2 + e^{\delta\beta} + e^{\delta\gamma}).$$

Putting all together, we obtain

$$\int \exp[\delta NF(\rho_N)] dW^{\otimes N}$$
$$\leq \exp[NT(2\delta - \delta(e^{-\beta} + e^{-\gamma}) - 2 + e^{\delta\beta} + e^{\delta\gamma})] \int \exp[NF_1(\rho_N)] dW^{\otimes N}$$
$$= \exp[NT(2\delta - \delta(e^{-\beta} + e^{-\gamma}) - 2 + e^{\delta\beta} + e^{\delta\gamma})],$$

from which the conclusion follows easily. □

COMPLETING THE PROOF OF PROPOSITION 3.1. It remains to show the upper and the lower bounds (9) and (8). We prove them separately; our main tool is the Varadhan Lemma in the version in [15], Lemmas 4.3.4 and 4.3.6.

We deal first with the upper bound (9). Take $r \geq \max(e^\beta, e^\gamma)$, so that the function $F_r$ in (35) is upper semicontinuous. Denote by $\mathcal{P}_N$ the distribution of $\rho_N$ under $P_N$, and by $\mathcal{W}_N$ its distribution under $W_r^{\otimes N}$. By (36)

(40) $$\frac{d\mathcal{P}_N}{d\mathcal{W}_N}(Q) = \exp[NF_r(Q)].$$

By Sanov's theorem (Theorem 6.2.10 in [15]), the sequence of probabilities $\mathcal{W}_N$ satisfies a large deviation principle with rate function $H(Q|W_r)$. Since $F_r$ is upper semicontinuous and satisfies the superexponential estimate in Lemma 5.7, we can apply Lemma 4.3.6 in [15], together with identity (39), to obtain the upper bound (9). The lower bound (8) is proved similarly, by taking $0 < r \leq \min(e^{-\beta}, e^{-\gamma})$, so that $F_r$ becomes lower semicontinuous, using (40) again and Lemma 4.3.4 in [15]. □

The remaining part of this section is devoted to the proof of Proposition 3.2. It mainly consists in giving an alternative representation of the rate function $I(Q)$.

Let now $Q \in \mathcal{M}_1(\mathcal{D}([0,T]) \times \mathcal{D}([0,T]))$. We associate with $Q$ the law of a time-inhomogeneous Markov process on $\{-1,1\}^2$ which evolves according to the following rules:

$$\sigma \to -\sigma$$

with intensity $e^{-\beta\sigma\omega}$,

$$\omega \to -\omega$$

with intensity $\exp\left[-\gamma\omega \sum_{\sigma,\tau \in \{-1,1\}} \sigma \Pi_t Q(\sigma,\tau)\right] = e^{-\gamma\omega m^\sigma_{\Pi_t Q}} = e^{-\gamma^Q_t \omega}$,

and with initial distribution $\lambda$. We denote by $P^Q$ the law of this process. In other words, $P^Q$ is the law of the Markov process on $\{-1,1\}^2$ with initial distribution $\lambda$ and time-dependent generator

$$\mathcal{L}^Q_t f(\sigma,\omega) = e^{-\beta\sigma\omega} \nabla^\sigma f(\sigma,\omega) + e^{-\gamma\omega m^\sigma_{\Pi_t Q}} \nabla^\omega f(\sigma,\omega).$$

LEMMA 5.8. *For every $Q \in \mathcal{M}_1(\mathcal{D}([0,T]) \times \mathcal{D}([0,T]))$ such that $I(Q) < +\infty$, we have*

$$I(Q) = H(Q|P^Q).$$

PROOF. We begin by observing that, since by assumption $I(Q) < \infty$, we have $H(Q|W) < +\infty$ and so by Lemma 5.4 it follows that $Q \in \mathcal{I}$, which implies that the integrals below are well defined. Using again Girsanov's formula for Markov chains in Proposition 5.1, we obtain

$$\int \log \frac{dP^Q}{dW}(\sigma[0,T], \omega[0,T]) \, dQ$$

$$= \int \left[ \int_0^T (1 - e^{-\beta\sigma(t)\omega(t)}) \, dt + \int_0^T (1 - e^{-\gamma\omega(t)\int \sigma \Pi_t Q(d\sigma,d\tau)}) \, dt \right.$$

$$+ \int_0^T (-\beta\sigma(t^-)\omega(t^-)) \, dN^\sigma_t$$

$$\left. + \int_0^T -\gamma\omega(t^-)\left[\int \sigma \Pi_{t^-} Q(d\sigma,d\tau)\right] dN^\omega_t \right] dQ$$

$$= \int \left[ \int_0^T (1 - e^{-\beta\sigma(t)\omega(t)}) \, dt + \int_0^T (1 - e^{-\gamma\omega(t)\int \sigma \Pi_t Q(d\sigma,d\tau)}) \, dt \right.$$

$$\left. + \beta \int_0^T \sigma(t)\omega(t) \, dN^\sigma_t + \gamma \int_0^T \omega(t)\left[\int \sigma \Pi_t Q(d\sigma,d\tau)\right] dN^\omega_t \right] dQ$$

$$= \int \left[ \int_0^T (1 - e^{-\beta\sigma(t)\omega(t)}) \, dt + \int_0^T (1 - e^{-\omega(t)\gamma^Q_t}) \, dt \right.$$

$$+ \beta \int_0^T \sigma(t)\omega(t)\,dN_t^\sigma + \int_0^T \omega(t)\gamma_{t-}^Q\,dN_t^\omega\bigg]dQ$$

$$= F(Q).$$

Finally, just observe that

$$I(Q) = \int dQ \log \frac{dQ}{dW} - \int dQ \log \frac{dP^Q}{dW}$$
$$= \int dQ \log \frac{dQ}{dP^Q} = H(Q|P^Q). \quad \Box$$

COMPLETING THE PROOF OF PROPOSITION 3.2. By properness of the relative entropy $[H(\mu|\nu) = 0 \Rightarrow \mu = \nu]$, from Lemma 5.8 we have that the equation $I(Q) = 0$ is equivalent to $Q = P^Q$. Suppose $Q^*$ is a solution of this last equation. Then, in particular, $q_t := \Pi_t Q^* = \Pi_t P^{Q^*}$. The marginals of a Markov process are solutions of the corresponding *forward equation* that, in this case, leads to the fact that $q_t$ is a solution of (10). This differential equation, being an equation in finite dimension with locally Lipschitz coefficients, has at most one solution in $[0,T]$. Since $P^{Q^*}$ is totally determined by the flow $q_t$, it follows that equation $Q = P^Q$ has at most one solution. The existence of a solution follows from the fact that $I(Q)$ is the rate function of a LDP, and therefore *must* have at least one zero, indeed, by (8) with $A = \mathcal{M}_1(\mathcal{D}[0,T] \times \mathcal{D}[0,T])$, we get $\inf_Q I(Q) = 0$. Since $I$ is lower semicontinuous, this inf is actually a minimum. $\Box$

5.2. *Proof of Theorem 3.4.* We first observe that the square $[-1,1]^2$ is stable for the flow of (14), since the vector field $V(x,y)$ points inward at the boundary of $[-1,1]^2$. It is also immediately seen that the equation $V(x,y) = 0$ holds if and only if $x = \tanh(\beta)\tanh(\gamma x)$ and $y = \frac{1}{\tanh(\beta)}x$. Moreover a simple convexity argument shows that $x = \tanh(\beta)\tanh(\gamma x)$ has $x = 0$ as unique solution for $\gamma \leq \frac{1}{\tanh(\beta)}$, while for $\gamma > \frac{1}{\tanh(\beta)}$ a strictly positive solution, and its opposite, bifurcate from the null solution. We have therefore found all equilibria of (14).

We now remark that (14) has no cycles (periodic solutions). Indeed, suppose $(x_t, y_t)$ is a cycle of period $T$. Then by the Divergence Theorem

$$(41) \qquad 0 \leq \int_0^T [V_1(x_t,y_t)\dot{x}_t + V_2(x_t,y_t)\dot{y}_t]\,dt = \int_C \operatorname{div} V(x,y)\,dx\,dy,$$

where $V_1, V_2$ are the components of $V$ and $C$ is the open set enclosed by the cycle. But a simple direct computation shows that $\operatorname{div} V(x,y) < 0$ in all of $[-1,1]^2$, so that (41) cannot hold.

It follows by the Poincaré–Bendixon theorem that every solution must converge to an equilibrium as $t \to +\infty$. This completes the proof of (i). The matrix of the linearized system is

$$DV(0,0) = \begin{pmatrix} -2\cosh(\beta) & 2\sinh(\beta) \\ 2\gamma & -2 \end{pmatrix}$$

from which also (ii) and (iii) are readily shown. It remains to show (iv). For $\gamma > \frac{1}{\tanh(\beta)}$, we let $v_s$ be an eigenvector of the negative eigenvalue of $DV(0,0)$. By the Stable Manifold Theorem (see Section 2.7 in [30]), the set of initial conditions that are asymptotically driven to $(0,0)$ form a one-dimensional manifold $\Gamma$ that is tangent to $v_s$ at $(0,0)$. Since any solution converges to an equilibrium point, and solutions starting in $\Gamma^c$ cannot cross $\Gamma$ (otherwise uniqueness would be violated), the remaining part of statement (iv) follows.

### 5.3. Proof of Theorem 3.6.

PROOF.  One key remark is the fact that the stochastic process $(m^\sigma_{\rho_N(t)}, m^\omega_{\rho_N(t)}, m^{\sigma\omega}_{\rho_N(t)})$ is a *sufficient statistic* for our model; in this context this means that its evolution is Markovian. This can be proved by checking that if we apply the generator $L$ in (3) to a function of the form $\varphi(m^\sigma_{\rho_N(t)}, m^\omega_{\rho_N(t)}, m^{\sigma\omega}_{\rho_N(t)})$, then we obtain again a function of $(m^\sigma_{\rho_N(t)}, m^\omega_{\rho_N(t)}, m^{\sigma\omega}_{\rho_N(t)})$. A long but straightforward computation actually gives

$$L\varphi(m^\sigma_{\rho_N(t)}, m^\omega_{\rho_N(t)}, m^{\sigma\omega}_{\rho_N(t)}) = [\mathcal{K}_N \varphi](m^\sigma_{\rho_N(t)}, m^\omega_{\rho_N(t)}, m^{\sigma\omega}_{\rho_N(t)}),$$

where

$$\begin{aligned}
\mathcal{K}_N \varphi(\xi, \eta, \theta) &= \frac{N}{4} \sum_{(j,k) \in \{-1,1\}^2} [j\xi + k\eta + jk\theta + 1] \\
&\quad \times \left\{ e^{-\beta jk} \left[ \varphi\left(\xi - \frac{2}{N}j, \eta, \theta - \frac{2}{N}jk\right) - \varphi(\xi, \eta, \theta) \right] \right. \\
&\quad\quad \left. + e^{-\gamma \xi k} \left[ \varphi\left(\xi, \eta - \frac{2}{N}k, \theta - \frac{2}{N}jk\right) - \varphi(\xi, \eta, \theta) \right] \right\}.
\end{aligned}$$

(42)

This implies that $\mathcal{K}_N$ is the infinitesimal generator of the three-dimensional Markov process $(m^\sigma_{\rho_N(t)}, m^\omega_{\rho_N(t)}, m^{\sigma\omega}_{\rho_N(t)})$. Note now that $(x_N(t), y_N(t), z_N(t))$ is obtained from $(m^\sigma_{\rho_N(t)}, m^\omega_{\rho_N(t)}, m^{\sigma\omega}_{\rho_N(t)})$ through a time dependent, linear invertible transformation. We call $T_t$ this transformation, that is,

$$T_t(\xi, \eta, \theta) = (\sqrt{N}(\xi - m^\sigma_t), \sqrt{N}(\eta - m^\omega_t), \sqrt{N}(\theta - m^{\sigma\omega}_t))$$

(the dependence on $N$ of $T_t$ is omitted in the notation). Therefore $(x_N(t), y_N(t), z_N(t))$ is itself a (time-inhomogeneous) Markov process, whose infinitesimal generator $\mathcal{H}_{N,t}$ can be obtained from (42) as follows:

$$\mathcal{H}_{N,t} f(x,y,z) = \mathcal{K}_N [f \circ T_t](T_t^{-1}(x,y,z)) + \frac{\partial}{\partial t}[f \circ T_t](T_t^{-1}(x,y,z)).$$

A simple computation gives then

$$\mathcal{H}_{N,t} f(x,y,z)$$
$$= \frac{N}{4} \sum_{(j,k) \in \{-1,1\}^2} \left[ j\frac{x}{\sqrt{N}} + k\frac{y}{\sqrt{N}} + jk\frac{z}{\sqrt{N}} + jm_t^\sigma + km_t^\omega + jkm_t^{\sigma\omega} + 1 \right]$$
$$\times \left\{ e^{-\beta jk} \left[ f\left(x - \frac{2}{\sqrt{N}} j, y, z - \frac{2}{\sqrt{N}} jk\right) - f(x,y,z) \right] \right.$$
(43)
$$\left. + e^{-\gamma(x/\sqrt{N} + m_t^\sigma)k} \left[ f\left(x, y - \frac{2}{\sqrt{N}} k, z - \frac{2}{\sqrt{N}} jk\right) - f(x,y,z) \right] \right\}$$
$$- \sqrt{N} \dot{m}_t^\sigma f_x(x,y,z) - \sqrt{N} \dot{m}_t^\omega f_y(x,y,z) - \sqrt{N} \dot{m}_t^{\sigma\omega} f_z(x,y,z),$$

where $f_x$ stands for $\frac{\partial f}{\partial x}$, and similarly for the other derivatives. At this point we compute the asymptotics of $\mathcal{H}_{N,t} f(x,y,z)$ as $N \to +\infty$, assuming $f : \mathbb{R}^3 \to \mathbb{R}$ a $\mathcal{C}^3$ function with compact support. First of all we make a Taylor expansion of terms like

(44)
$$f\left(x - \frac{2}{\sqrt{N}} j, y, z - \frac{2}{\sqrt{N}} jk\right) - f(x,y,z)$$
$$= -\frac{2}{\sqrt{N}} f_x(x,y,z) - \frac{2}{\sqrt{N}} f_z(x,y,z)$$
$$+ \frac{2}{N} f_{xx}(x,y,z) + \frac{2}{N} f_{zz}(x,y,z) + \frac{4}{N} f_{xz}(x,y,z) + o\left(\frac{1}{N}\right)$$

and

(45)
$$e^{-\gamma(x/\sqrt{N})} = 1 - \gamma\left(\frac{x}{\sqrt{N}}\right) + o\left(\frac{1}{\sqrt{N}}\right).$$

Note that, since all derivatives of $f$ are bounded, the remainder in (44) is $o(\frac{1}{N})$ *uniformly* in $(x,y,z) \in \mathbb{R}^3$. Moreover, the remainder in (45) is $o(\frac{1}{\sqrt{N}})$ uniformly for $x$ in a compact set. Therefore, since $f$ has compact support, when we use (44) and (45) to replace the corresponding terms in (43), we obtain remainders whose bounds are uniform in $\mathbb{R}^3$. When (44) and (45) are plugged into (43), all terms of order $\sqrt{N}$ coming from the sum over $(j,k) \in$ $\{-1,1\}^2$ are canceled by the terms $\sqrt{N}\dot{m}_t^\sigma f_x(x,y,z) - \sqrt{N}\dot{m}_t^\omega f_y(x,y,z) - \sqrt{N}\dot{m}_t^{\sigma\omega} f_z(x,y,z)$. It follows then by a straightforward computation that

$$\lim_{N\to\infty} \sup_{t\in[0,T]} \sup_{x,y,z\in\mathbb{R}^3} |\mathcal{H}_{N,t}f(x,y,z) - \mathcal{H}_t f(x,y,z)| = 0,$$

where

$$\begin{aligned}
\mathcal{H}_t f(x,y,z) = 2\{&f_x[-x\cosh(\beta) + y\sinh(\beta)] \\
&+ f_y[-\gamma x m_t^\omega \sinh(\gamma m_t^\sigma) + \gamma x \cosh(\gamma m_t^\sigma) - y\cosh(\gamma m_t^\sigma)] \\
&+ f_z[x\sinh(\gamma m_t^\sigma) + \gamma x m_t^\sigma \cosh(\gamma m_t^\sigma) \\
&\qquad - \gamma x m_t^{\sigma\omega} \sinh(\gamma m_t^\sigma) - z\cosh(\beta) - z\cosh(\gamma m_t^\sigma)] \\
&+ f_{xx}[-m_t^{\sigma\omega}\sinh(\beta) + \cosh(\beta)] \\
&+ f_{yy}[-m_t^\omega \sinh(\gamma m_t^\sigma) + \cosh(\gamma m_t^\sigma)] \\
&+ f_{zz}[-m_t^{\sigma\omega}\sinh(\beta) + \cosh(\beta) \\
&\qquad - m_t^\omega \sinh(\gamma m_t^\sigma) + \cosh(\gamma m_t^\sigma)] \\
&+ 2f_{xz}[-m_t^\sigma \sinh(\beta) + m_t^\omega \cosh(\beta)] \\
&+ 2f_{yz}[m_t^\sigma \cosh(\gamma m_t^\sigma) - m_t^{\sigma\omega} \sinh(\gamma m_t^\sigma)]\}
\end{aligned} \qquad (46)$$

is the infinitesimal generator of the linear diffusion process (18). Using Theorem 1.6.1 in [17], the proof is completed if we show that $(x_N(0), y_N(0), z_N(0))$ converges as $N \to +\infty$, in distribution to $(x(0), y(0), z(0))$. This last statement follows by the standard Central Limit Theorem for i.i.d. random variables; indeed, by assumption, $(\sigma_i(0), \omega_i(0))$ are independent with law $\lambda$, and (19) is just the covariance matrix under $\lambda$ of $(\sigma(0), \omega(0), \sigma(0)\omega(0))$. It should be pointed out that Theorem 1.6.1 in [17] does not deal explicitly with time-dependent generators, as is the case here. To fix this point it is enough to introduce an additional variable, $\tau(t) := t$, and consider the process $\alpha(t) := (x(t), y(t), z(t), \tau(t))$, whose generator is time-homogeneous. This argument, together with the fact that the convergence of $\mathcal{H}_{N,t}f(x,y,z)$ to $\mathcal{H}_t f(x,y,z)$ is uniform in both $(x,y,z)$ and $t$, completes the proof. □

5.4. *Proof of Theorem 4.1.* We start with a technical lemma.

LEMMA 5.9. *For $t \in [0,T]$ we have the convergence in distribution*

$$\sqrt{N}\left(\frac{\sum_j l_{\sigma_j(t)}}{N} - L(t)\right) \to X \sim N\left(0, \frac{(l_1 - l_{-1})^2 V(t)}{4}\right),$$

*where $L(t)$ is defined in (28) and $V(t)$ in (24).*

PROOF. Define, for $x \in \{-1, 1\}$, the quantity $A_x^N(t)$ as the number of $\sigma_i$ that, at a given time $t$, are equal to $x$. We may then write $\frac{1+m_N^\sigma(t)}{2} = \frac{A_1^N(t)}{N}$ and $\frac{1-m_N^\sigma(t)}{2} = \frac{A_{-1}^N(t)}{N}$. Recall moreover that for $N \to \infty$, $m_N^\sigma(t) \to m_t^\sigma$. We then have

$$\sqrt{N}\left(\frac{\sum_j l_{\sigma_j(t)}}{N} - L(t)\right)$$

$$= \sqrt{N}\left(\frac{l_1 A_1^N(t) + l_{-1} A_{-1}^N(t)}{N} - L(t)\right)$$

$$= \sqrt{N}\left(l_1 \frac{1+m_N^\sigma(t)}{2} + l_{-1} \frac{1-m_N^\sigma(t)}{2} - L(t)\right)$$

$$= \sqrt{N}\left(\frac{(l_1+l_{-1})}{2} + \frac{(l_1-l_{-1})}{2}m_N^\sigma(t) - \frac{(l_1-l_{-1})}{2}m_t^\sigma - \frac{(l_1+l_{-1})}{2}\right)$$

$$= \sqrt{N}\left(\frac{(l_1-l_{-1})}{2}(m_N^\sigma(t) - m_t^\sigma)\right) \to X \sim N\left(0, \frac{(l_1-l_{-1})^2 V(t)}{4}\right),$$

where the last convergence follows from Corollary 3.7 noticing that $m_N^\sigma(t) = m_{\rho_N(t)}^\sigma$. $\square$

PROOF OF THEOREM 4.1. We have to check that

$$\sqrt{N}\left(\frac{L^N(t)}{N} - L(t)\right) \to Y \sim N(0, \hat{V}(t)),$$

where $\hat{V}(t)$ is defined in (29).

Separating the firms according to whether their $\sigma_j(t)$ is $+1$ or $-1$,

$$\sqrt{N}\left(\frac{\sum_j L_j(t)}{N} - L(t)\right) = \sqrt{N}\left(\frac{\sum_{j:\sigma_j(t)=1} L_j(t) + \sum_{j:\sigma_j(t)=-1} L_j(t)}{N} - L(t)\right).$$

We then add and subtract $\sum_j l_{\sigma_j(t)}$ to obtain

(47)
$$\sqrt{N}\left(\frac{\sum_{j:\sigma_j(t)=1}(L_j(t) - l_1)}{N} + \frac{\sum_{j:\sigma_j(t)=-1}(L_j(t) - l_{-1})}{N} + \frac{\sum_j l_{\sigma_j(t)}}{N} - L(t)\right).$$

Since we have only independence conditionally on $\underline{\sigma}(t)$, we need to check whether the CLT still applies. Let us show the convergence of the corresponding characteristic functions:

$$E\left[\exp\left\{ir\frac{L^N(t) - NL(t)}{\sqrt{N}}\right\}\right]$$

(48)
$$= E\left[E\left[\exp\left\{ir\left(\frac{\sum_{j:\sigma_j(t)=1}(L_j(t) - l_1)}{\sqrt{N}}\right.\right.\right.\right.$$
$$+ \frac{\sum_{j:\sigma_j(t)=-1}(L_j(t) - l_{-1})}{\sqrt{N}}$$
$$\left.\left.\left.\left.+ \frac{\sum_j l_{\sigma_j(t)} - NL(t)}{\sqrt{N}}\right)\right\}\bigg|\underline{\sigma}(t)\right]\right].$$

The last of the three terms is measurable with respect to the sigma algebra generated by $\sigma(t)$ so that we can take it out from the inner expectation. Because of the conditional independence we can separate the remaining terms in the product of conditional expectations:

$$E\left[\exp\left\{ir\frac{\sum_{j:\sigma_j(t)=1}(L_j(t) - l_1)}{\sqrt{N}}\right\}\bigg|\underline{\sigma}(t)\right]$$
$$\times E\left[\exp\left\{ir\frac{\sum_{j:\sigma_j(t)=-1}(L_j(t) - l_{-1})}{\sqrt{N}}\right\}\bigg|\underline{\sigma}(t)\right].$$

By conditional independence,

$$E\left[\exp\left\{ir\frac{\sum_{j:\sigma_j(t)=1}(L_j(t) - l_1)}{\sqrt{N}}\right\}\bigg|\underline{\sigma}(t)\right]$$
$$= \prod_{j=1}^{A_1^N(t)} E\left[\exp\left\{ir\frac{L_j(t) - l_1}{\sqrt{N}}\right\}\bigg|\underline{\sigma}(t)\right] = \left[1 - \frac{v_1}{2}\frac{r^2}{N} + o\left(\frac{1}{N}\right)\right]^{A_1^N(t)},$$

where the last equality follows because $l_1$ and $v_1$ are the first two conditional moments of $L_j(t)$.

Recalling that $\frac{A_1^N(t)}{N} = \frac{1+m_N^\sigma(t)}{2}$ converges almost surely to $\frac{1+m_t^\sigma}{2}$ we have that

$$\lim_{N\to\infty}\left[1 - \frac{v_1}{2}\frac{r^2}{N} + o\left(\frac{1}{N}\right)\right]^{A_1^N(t)} = \lim_{N\to\infty}\left[1 - \frac{v_1}{2}\frac{r^2}{A_1^N(t)}\frac{A_1^N(t)}{N} + o\left(\frac{1}{N}\right)\right]^{A_1^N(t)}$$
$$= \exp\left[-\frac{r^2}{2}\frac{1+m_t^\sigma}{2}v_1\right].$$

The same argument holds for the terms where $\sigma_j(t) = -1$. Since $\frac{A_{-1}^N(t)}{N} \to \frac{1-m_t^\sigma}{2}$, we have

$$\lim_{N\to\infty}\left[1 - \frac{v_{-1}}{2}\frac{r^2}{A_{-1}^N(t)}\frac{A_{-1}^N(t)}{N} + o\left(\frac{1}{N}\right)\right]^{A_{-1}^N(t)} = \exp\left[-\frac{r^2}{2}\frac{1-m_t^\sigma}{2}v_{-1}\right].$$

Finally, recall from Lemma 5.9 that $\frac{\sum_j l_{\sigma_j(t)} - NL(t)}{\sqrt{N}}$ converges to $X \sim N(0, \frac{(l_1-l_{-1})^2 V(t)}{4})$, so that

$$\lim_{N\to\infty} E\left[\exp\left\{ir\frac{\sum_j l_{\sigma_j(t)} - NL(t)}{\sqrt{N}}\right\}\right] = \exp\left[-\frac{r^2}{2}\frac{(l_1-l_{-1})^2 V(t)}{4}\right].$$

Thus, denoting by $E[\cdots|\sigma(t)]$ the inner conditional expectation in (48), we have shown that

$$\lim_{N\to\infty} E[\cdots|\sigma(t)] = \exp\left[-\frac{r^2}{2}\frac{(l_1-l_{-1})^2 V(t)}{4}\right]\exp\left[-\frac{r^2}{2}\frac{1+m_t^\sigma}{2}v_1\right]$$

$$\times \exp\left[-\frac{r^2}{2}\frac{1-m_t^\sigma}{2}v_{-1}\right]$$

$$= \exp\left[-\frac{r^2}{2}\hat{V}(t)\right].$$

By the Dominated Convergence Theorem, taking the limit as $N \to +\infty$ in (48), we can interchange the limit with the outer expectation, and the proof is completed. $\square$

**6. Conclusions and possible extensions.** In this paper we have described propagation of *financial distress* in a network of firms linked by business relationships.

We have proposed a model for *credit contagion*, based on interacting particle systems, and we have quantified the impact of contagion on the losses suffered by a financial institution holding a large portfolio with positions issued by the firms.

Compared to the existing literature on credit contagion, we have proposed a *dynamic* model where it is possible to describe the evolution of the indicators of financial distress. In this way we are able to compute the distribution of the losses in a large portfolio for any time horizon $T$, via a suitable version of the central limit theorem.

The peculiarity of our model is the fact that the changes in rating class (the $\sigma$ variables) are related to the degree of health of the system (the global indicator $m^\sigma$). There is a further characteristic of the firms that is summarized by a second variable $\omega$ (a liquidity indicator) and that describes the ability of the firm to act as a buffer against adverse news coming from the market. The evolution of the pair $(\sigma, \omega)$ depends on two parameters $\beta$ and $\gamma$, which indicate the strength of the interaction.

The fact that our model leads to endogenous financial indicators that describe the general health of the systems has allowed us to view a credit crisis as a microeconomic phenomenon. This has also been exemplified through simulation results.

The model we have proposed in this paper exhibits some phenomena having interesting financial interpretation. There are many extensions that could make the model more flexible and realistic, allowing also calibration to real data. One of them, concerning the symmetry of the model, has already been mentioned in Remark 2.1. Other more substantial extensions are the following:

- In real applications, the variable $\sigma$ denoting the rating class is not binary; one could extend the model by taking $\sigma$ to be valued in a finite, totally ordered set.
- One could assume the fundamental values $\omega_i$ to be $\mathbb{R}^+$-valued, and evolving according to the stochastic differential equation

$$d\omega_i(t) = \omega_i(t)[f(m_N^\sigma(t))\,dt + g(m_N^\sigma(t))\,dB_i(t)] + dJ_i(t),$$

where $f$ and $g$ are given functions, the $B_i(\cdot)$ are independent Brownian motions, and $J_i(\cdot)$ is a pure jump process whose intensity is a function of $\omega_i(t)$ and $m_N^\sigma(t)$.
- An interesting extension of the above model consists in letting the functions $a(\cdot,\cdot,\cdot)$ and $b(\cdot,\cdot,\cdot)$ in (1) be random rather than deterministic; in particular they may depend on (possibly time-dependent) exogenous macroeconomic variables.
- The mean-field assumption may be weakened by assuming that the rate at which $\omega_i$ changes depends on an $i$-dependent weighted global health of the form

$$m_{N,i}^\sigma := \frac{1}{N}\sum_{j=1}^N J\left(\frac{i}{N},\frac{j}{N}\right)\sigma_j,$$

where $J:[0,1]^2 \to \mathbb{R}$ is a function describing the interaction between pairs of firms. In other words, the $i$th firm "feels" the information given by the rating of the other firms in a nonuniform way.

Other generalizations could be useful, in particular to introduce inhomogeneity in the model. In principle, the extensions listed above could be treated by the same techniques used in this paper.

## APPENDIX: THE EIGENVALUES OF THE MATRIX $A$ IN THEOREM 3.6

We begin by writing down explicitly the limit matrix $A$:

$$A = 2\begin{pmatrix} -\cosh(\beta) \\ -\gamma\dfrac{\sinh(\gamma m_*^\sigma)}{\cosh(\gamma m_*^\sigma)}\sinh(\gamma m_*^\sigma) + \gamma\cosh(\gamma m_*^\sigma) \\ \sinh(\gamma m_*^\sigma) + \gamma m_*^\sigma \cosh(\gamma m_*^\sigma) + \gamma\dfrac{\sinh(\beta) + m_*^\sigma \sinh(\gamma m_*^\sigma)}{\cosh(\beta) + \cosh(\gamma m_*^\sigma)}\sinh(\gamma m_*^\sigma) \end{pmatrix}$$

$$\begin{matrix} \sinh(\beta) & 0 \\ -\cosh(\gamma m_*^\sigma) & 0 \\ 0 & -(\cosh(\beta)+\cosh(\gamma m_*^\sigma)) \end{matrix}\Bigg)$$

where for the first term in the second row we have used (16). By direct computation, one shows that the eigenvalues of $A$ are given by the following expressions:

$$\lambda_1 = -2(\cosh(\beta)+\cosh(\gamma m_*^\sigma)),$$

$$\lambda_2 = -\bigg\{\cosh(\beta)+\cosh(\gamma m_*^\sigma)$$

(49)
$$+\sqrt{(\cosh(\beta)-\cosh(\gamma m_*^\sigma))^2 + 4\gamma\frac{\sinh(\beta)}{\cosh(\gamma m_*^\sigma)}}\bigg\},$$

$$\lambda_3 = -\bigg\{\cosh(\beta)+\cosh(\gamma m_*^\sigma)$$

$$-\sqrt{(\cosh(\beta)-\cosh(\gamma m_*^\sigma))^2 + 4\gamma\frac{\sinh(\beta)}{\cosh(\gamma m_*^\sigma)}}\bigg\}.$$

Note that these eigenvalues are all real, and that clearly $\lambda_1, \lambda_2 < 0$. Moreover, $\lambda_3 < 0$ if and only if

(50) $$\frac{\gamma}{\gamma_c} < \cosh^2(\gamma m_*^\sigma)$$

where $\gamma_c = \frac{1}{\tanh(\beta)}$.

(a) If $\gamma < \gamma_c$, then by part (i) in Theorem 3.4 we have $m_*^\sigma = 0$. In this case (50) holds, because

$$\frac{\gamma}{\gamma_c} < 1 = \cosh^2(\gamma \cdot 0).$$

In this case the matrix $A$ has three different real eigenvalues, all strictly negative.

(b) If $\gamma = \gamma_c$, we still have $m_*^\sigma = 0$, but it is immediately seen that $\lambda_3 = 0$.

(c) Finally, if $\gamma > \gamma_c$, set $y = \gamma m_*^\sigma$; by (15) we have

(51) $$m_*^\sigma = \frac{1}{\gamma_c}\tanh(\gamma m_*^\sigma) \Leftrightarrow y = \frac{\gamma}{\gamma_c}\tanh(y).$$

Then (50) is equivalent to showing that

(52) $$\frac{\gamma}{\gamma_c} < \cosh^2(y)$$

and from (51) we obtain

$$\frac{\gamma}{\gamma_c} = \frac{y}{\tanh(y)} = \frac{y}{\sinh(y)} \cosh(y) < \cosh(y) < \cosh^2(y)$$

because $y/\sinh(y) < 1$ and $\cosh(y) < \cosh^2(y)$, since $y = \gamma m_*^\sigma > 0$ if $\gamma > \gamma_c$. Then, in this case too, the matrix $A$ has three different real eigenvalues, all strictly negative.

**Acknowledgment.** The authors would like to acknowledge the extremely careful reading of the paper and the useful suggestions made by an anonymous referee.

P. Dai Pra
W. J. Runggaldier
E. Sartori
Dipartimento di Matematica Pura ed Applicata
University of Padova
63, Via Trieste
I-35121-Padova
Italy
E-mail: daipra@math.unipd.it
       runggal@math.unipd.it
       esartori@math.unipd.it

M. Tolotti
Istituto di Metodi Quantitativi
Bocconi University
25, Via Sarfatti
I-20136 Milano
Italy
and
Scuola Normale Superiore
Pisa
Italy
E-mail: tolotti@unibocconi.it
and
Department of Applied Mathematics
University of Venice
Venice
Italy
E-mail: tolotti@unive.it